\documentclass{aastex63}

\submitjournal{ApJ}
\shorttitle{Globular cluster iron spreads: II. Protocluster Metallicities}
\shortauthors{Bailin \&\ von Klar}

\newcommand{\feh}{{\ensuremath{\mathrm{[Fe/H]}}}}
\newcommand{\fehinit}{{\ensuremath{\mathrm{[Fe/H]}_{\mathrm{init}}}}}
\newcommand{\Minit}{{\ensuremath{M_{\mathrm{init}}}}}

\defcitealias{Bailin18}{B18}
\defcitealias{Bailin19}{B19}
\defcitealias{Meszaros20}{M20}

\begin{document}
\title{Globular Cluster Intrinsic Iron Abundance Spreads: II. Protocluster Metallicities and the Age-Metallicity Relations of Milky Way Progenitors}

\email{jbailin@ua.edu}
\author[0000-0001-6380-010X]{Jeremy Bailin}
\affiliation{Department of Physics and Astronomy, University of Alabama Box 870324, Tuscaloosa, AL, 35487-0324, USA}

\author{Ryker von Klar}
\affiliation{Department of Physics and Astronomy, University of Alabama Box 870324, Tuscaloosa, AL, 35487-0324, USA}

\correspondingauthor{Jeremy Bailin}



\begin{abstract}
Intrinsic iron abundance spreads in globular clusters, although usually small, are very common, and are signatures of self enrichment: some stars within the cluster have been enriched by supernova ejecta from other stars within the same cluster. We use the Bailin (2018) self enrichment model to predict the relationship between properties of the protocluster --- its mass and the metallicity of the protocluster gas cloud --- and the final observable properties today --- its current metallicity and the internal iron abundance spread. We apply this model to an updated catalog of Milky Way globular clusters where the initial mass and/or the iron abundance spread is known to reconstruct their initial metallicities. We find that with the exception of the known anomalous bulge cluster Terzan~5 and three clusters strongly suspected to be nuclear star clusters from stripped dwarf galaxies, the model provides a good lens for understanding their iron spreads and initial metallicities. We then use these initial metallicities to construct age-metallicity relations for kinematically-identified major accretion events in the Milky Way's history. We find that using the initial metallicity instead of the current metallicity does not alter the overall picture of the Milky Way's history, since the difference is usually small, but does provide information that can help distinguish which accretion event some individual globular clusters with ambiguous kinematics should be associated with, and points to potential complexity within the accretion events themselves.
\end{abstract}



\section{Introduction}

In the current paradigm of galaxy formation, galaxies form in dark matter halos that grow over time by accreting surrounding material due to gravitational instability \citep{WhiteRees78}. Some of this accretion is smooth, while some is in the form of clumps that have themselves already turned into halos with their own galaxies; galaxy formation is thus a hierarchical process where galaxies form and grow and merge to form larger structures. A key to understanding the history of galaxies, and particularly the Milky Way, is thus to understand this accretion history.

Globular clusters (GCs) are a powerful probe of this process \citep[e.g.][]{SearleZinn78}.
Galaxies build up their metal abundance over time as heavy elements are released by nucleosynthesis in stars and supernovae into the surrounding gas. The gas metallicity therefore increases with time, and stars that form from that gas are usually expected to preserve that metallicity, resulting in an age-metallicity relation of the stars \citep[e.g.][]{Tinsley80,Edvardsson93,Piatti12,Haywood13,Snaith16}. This is particularly relevant for GCs because their ages and metallicities can be well determined: the GCs belonging to a particular galaxy are expected to exhibit an age-metallicity relation today that reflects the enrichment history of the galaxy in which they formed. They can therefore constrain the merger history of the Milky Way: the GCs from each accreted galaxy should follow a distinct age-metallicity relation that we can see today. This type of analysis has revealed that the Milky Way's GCs split into a metal-rich branch associated with GCs formed within the main Milky Way progenitor, and a metal-poor branch associated with accreted galaxies \citep{MarinFranch09,ForbesBridges10,Leaman13}.

More detailed histories have been obtained by \citet{Kruijssen19-catalog,Kruijssen20} by comparing an amalgamated catalog of 96 Milky Way GCs that have good age and metallicity measurements to E-MOSAICS, a series of hydrodynamic galaxy formation simulations that simultaneously follow the formation and evolution of GCs \citep{Pfeffer18,Kruijssen19-emosaics}. Their analysis finds 5 main satellite accretion events: in chronological order of accretion time they are Kraken, the Helmi streams, Sequoia, Gaia-Enceladus, and Sagittarius, with well constrained accretion redshifts and galaxy masses. These events, identified by the ages and metallicities of GCs, mostly match streams and phase space structures that have been kinematically identified using data from the Gaia and H3 surveys \citep{Massari19,Yuan20,Naidu20,Horta20}, although there is some ambiguity about the progenitor assignment for some individual GCs.

The discussion so far has been predicated on the idea that GCs are indeed fossils whose chemical abundances trace the conditions under which they formed. However, this is an oversimplification. The abundance patterns of GCs show unambiguous evidence of multiple populations whose abundances of intermediate elements such as oxygen and sodium have been dramatically influenced by self enrichment by the nucleosynthetic products of stars \textit{within the cluster itself} \citep[e.g.][]{Carretta09,Piotto15}. In other words, there is a difference between the abundances of the stars in the cluster today, which is what we can measure, and those of the gas from which they formed, which is what we want to know.

Although the existence of multiple populations is well established, a single model that explains the observations is not. The proposed nucleosynthetic sites of the intermediate mass elements, such as asymptotic giant branch (AGB) stars, very massive stars, or fast rotating massive stars, all conflict strongly with observations \citep[see][]{BastianLardo18}. It seems likely that multiple populations within GCs are not a single phenomenon, but the manifestation of several physical processes.

In this context, iron provides an intriguing starting point. Although iron abundance spreads within GCs are smaller than those of many other elements, most GCs do have measurable non-zero spreads \citep{WillmanStrader12,Leaman13,Renzini13,Bailin19}, implying that iron also experiences self enrichment; this is also suggested by the existence of the mass-metallicity relation of GCs within massive galaxies (the ``blue tilt''; \citealp{Harris06,Mieske06}). Unlike lower mass elements, which can be produced in a variety of types of stars, iron is only produced in supernovae. Therefore, we can attempt to understand the role that supernovae play in the self enrichment of iron in GCs much more cleanly than we can understand self enrichment of other elements.

In other words, if we want to use the metallicities of GCs to infer the Milky Way's accretion history, iron serves simultaneously as the main tracer of the overall metallicity, and also one where we might most easily be able to correct for the effects of self enrichment.

In \citet{Bailin18} (hereafter \citetalias{Bailin18}), we presented a model for supernova self enrichment of GCs based on the clumpiness of observed star formation events in the Galaxy. This model predicts the amount of self enrichment of iron a cluster has experienced, and therefore can be used to correct for it. In this paper, we present predictions of that model for a broad grid of protocluster properties (Section~\ref{sec:model}), compare the model to an updated observational catalog (Section~\ref{sec:obsdata}), use the model to infer the initial metallicites of the star formation events in which Milky Way GCs formed (Section~\ref{sec:fehinit}), and see what effect this correction has on the assignment of GCs to specific accretion events (Section~\ref{sec:age-metal}). Conclusions are presented in Section~\ref{sec:conclusions}.

\section{Self-enrichment model}\label{sec:model}

In \citetalias{Bailin18}, we developed a phenomenological model for how clumpiness in star forming regions could result in iron abundance spreads.
There is other recent theoretical work aimed at understanding self enrichment in GCs from a variety of perspectives that range in complexity and generalizability. For example, \citet{Jimenez21} present a more detailed phenomenological model that comes to similar qualitative conclusions as us. \citet{Wirth21} use an analytic analysis to argue that star formation in GCs must have been extended and had a supernova ejecta retention fraction similar to that produced by our model, while \citet{Marino21} use a similar analysis and find a clear correlation between retention fraction and the cluster mass, in contrast to \citet{Wirth21} who find none. Hydrodynamic simulations of GC formation in a cosmological context \citep[e.g.][]{Li19,Ma20} or specifically aimed at particular systems with the most extreme abundance spreads \citep{Bekki19} are also capable of producing systems in which self enrichment occurs and result in multiple populations.

In \citetalias{Bailin18}, we used as an example a gas cloud that had an initial metallicity of $\feh = -1.95$. In this section, we demonstrate how the initial metallicity and the initial cluster mass together influence the iron abundance spread.

\subsection{Model Description}

Here we give a brief overview of the GCZCSE model, which is described in detail in B18. Each GC is envisioned to form in an initially well-mixed self-gravitating protocluster gas cloud with iron abundance \fehinit. Each cloud fragments into a distinct number of clumps whose number scales $\propto M^{0.60}$. Each clump undergoes a single star formation event during which $30\%$ of its gas turns into stars, with stars formed according to a power-law initial mass function with cutoffs at low and high mass to give the correct number of core collapse supernovae per unit mass of stars formed according to a \citet{Chabrier03} initial mass function (IMF); see \citet{BH09}. The clumps within a cloud have their star formation events spread out over a timescale $\sigma_{\mathrm{tform}} \approx 10^7$~yr.

The newly-formed stellar population within each clump produces metal-rich ejecta and energy via core collapse supernovae. Type Ia supernovae are not included because the number of prompt SNIae is small: even assuming that the GCZCSE self enrichment extends over 50~Myr, and that SNIae begin immediately once the first stars form (before the formation of the first white dwarfs), there would only be $\sim 5$ over the relevant time period in a $10^6~M_{\odot}$ cluster, compared to $\sim 10^4$ core collapse supernovae \citep{Strolger20}. Therefore, even though each Type Ia supernova produces more iron, the combined effect of them is negligible compared to the core collapse supernovae at these times.

The supernova ejecta mixes with the surrounding material, and then a fraction of the newly-produced metals are lost; the fraction depends on the energy balance between the depth of the gravitational potential well and the energy injected by supernovae, and also on the mixing efficiency. The newly-produced metals that remain enhance the metallicity of the remaining gas cloud, so clumps that have later star formation events become enriched by star formation in earlier-forming clumps, resulting in both a net increase in the mean metallicity of the total stellar population, and a spread in the metallicity of stars from different clumps. The parameters of the model have been calibrated based on observations of star forming regions. For full details, see \citetalias{Bailin18}.

\subsection{Model Predictions}\label{sec:modelpred}

\begin{figure*}
\plotone{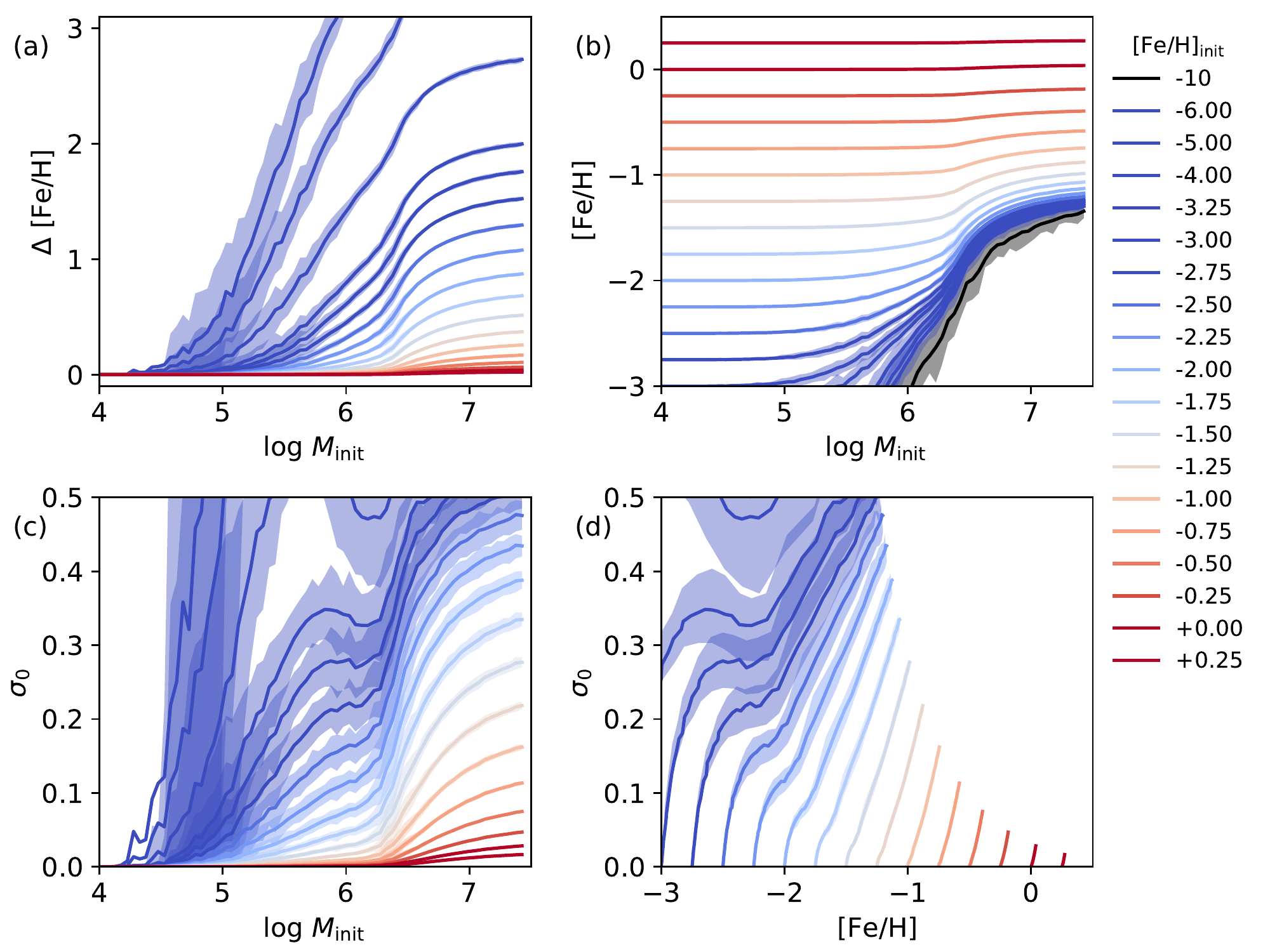}
\caption{\label{fig:FeH_model_tracks}%
Predictions of the \citetalias{Bailin18} GC self enrichment model for different values of the initial metallicity, \fehinit. In each panel, the solid lines denote the mean trend, while the shaded region encloses 68\%\ of the model instantiations and gives a measurement of the theoretical uncertainty due to the inherent stochasticity of the IMF. Colors denote \fehinit; the black line indicates the predictions for a virtually-pristine gas cloud.
 (a) The difference between the mean metallicities of the stars in the GC today compared to the initial gas metallicity, as a function of the initial mass of stars that form in the cluster, \Minit. Note that the present-day GC's mass is less than \Minit\ due to stellar evolution and tidal stripping. (b) The final mean iron abundance of the stars in a GC as a function of initial GC mass. (c) Internal spread in the iron abundances of the stars in a GC, $\sigma_0$, as a function of initial GC mass. (d) Internal spread in the iron abundances of the stars in a GC as a function of their final mean iron abundance.%
}
\end{figure*}

Figure~\ref{fig:FeH_model_tracks} shows the predictions of GCZCSE for the final iron abundance, \feh, the amount that the mean iron abundance increases from the initial value, $\Delta \feh \equiv \feh - \fehinit$, and the internal spread in iron abundance, $\sigma_0$, as a function of the initial metallicity, \fehinit, and initial cluster stellar mass, \Minit. $\sigma_0$ is defined as the $1\sigma$ standard deviation of the stellar mass-weighted \feh\ distribution of stars formed in the cluster.
Because the model is stochastic, different realizations of the same parameters sample the IMF differently and therefore end up with somewhat different final values. Each set of parameters was sampled 100 times with different random seeds in order to quantify this. In panels (a) through (c), the 16th-to-84th percentile range of model outputs is shown as the vertical extent of the shaded region around the mean model track. For panel (d), this would not give an accurate measurement because the quantities are correlated between different realizations (e.g. at low mass, there is quite a bit of stochastic uncertainty in each of $\Delta \feh$ and $\sigma_0$, as seen at the left side of panels (a) and (c), but they are strongly correlated: realizations with high $\Delta \feh$ also have high $\sigma_0$, so the effect of stochasticity does not introduce uncertainty to the relation between \feh\ and $\sigma_0$), so instead we calculated the width in this panel by finding the model realizations whose final \feh\ and $\sigma_0$ lay closest to each point along the mean relation and calculating their 16th-to-84th percentile range in the direction perpendicular to the mean relation at that point.

Note that \Minit\ is the total mass of stars initially produced. Mass loss due to stellar evolution and tidal evolution of the cluster within the Milky Way environment mean that the currently-observed mass of the cluster is lower than \Minit. A number of important conclusions stand out in panels (a) through (c):
\begin{enumerate}
 \item More massive GCs have more self enrichment. Although the details depend on the spatial distribution of the cloud, this is a basic consequence of the energy balance between the kinetic energy produced by supernovae, which scales as $M$, and the gravitational potential energy, which scales as $M^2$.
 \item Self-enrichment has a more profound effect on low-metallicity GCs. This is because  a given stellar population produces the same absolute amount of metals, but it has a much larger relative effect if the existing metallicity is low.
 \item The stochasticity of the GCZCSE model is particularly important at low metallicity, where the ejecta of each individual supernova can have a significant effect. The stochastic effects on each of \feh\ and $\sigma_0$ are largest at low mass, where the IMF is more sparsely sampled, but due to way these values are correlated the effects of stochasticity in the \feh-$\sigma_0$ plane are largest at large mass.
\item At low mass, \fehinit\ and \feh\ match, but the increased importance of self-enrichment is apparent at high masses. For low-\fehinit\ high-\Minit\ models, there is a pileup of model tracks in panel (b) -- these clusters have very similar mean metallicities today regardless of what their initial gas cloud looked like. This degeneracy arises because, in this regime, the metals coming from self-enrichment, which depends only on cluster mass, dominates over the pre-existing metals. This is demonstrated by the black line, which shows the predictions for a pristine gas cloud, where \textbf{all} of the metals come from self-enrichment. Below this curve is a forbidden region -- if the self-enrichment model is correct, then \textbf{any high-mass GC should have a minimum metallicity}. The formation of a high-mass GC should inexorably lead to minimum metal content. If any observed GCs are found to confidently inhabit this region, it would be strong evidence against this model.
 \item The spread $\sigma_0$ is driven by the range between the lowest-metallicity stars and the highest-metallicity stars, which both increase in metallicity as the importance of self-enrichment increases. For extremely low-metallicity clusters, there is an interesting mass regime around $10^6~M_{\odot}$ where the lower bound increases faster than the upper bound, resulting in $\sigma_0$ plateauing or even decreasing with \Minit, even as the importance of self-enrichment is overall increasing. This set of circumstances is fairly sensitive to the details of how the physical prescriptions are implemented in the model, and so this predicted ``bump'' in $\sigma_0$ should not be overinterpreted.
\end{enumerate}

\label{sec:modeldiscussion}

Panel (d) takes \Minit\ out of the equation, and shows how \feh\ and $\sigma_0$ vary together, for models with the same \fehinit. Each track corresponds to the present-day observable properties of GCs that came from gas of the same \fehinit, with different tracks corresponding to different initial gas metallicities. \citetalias{Bailin18} demonstrated that because both $\Delta \feh$ and $\sigma_0$ are consequences of how much self-enrichment there is, regardless of \textit{why} the cluster ended up with that amount of self-enrichment, this plot is much more robust to the parameters of the model than the plots in the other panels. We find the following:
\begin{enumerate}
 \item The final properties of GCs produced by the same metal-poor gas can look profoundly different, with a range in [Fe/H] of over 1.5~dex and in $\sigma_0$ of nearly 0.5~dex. At higher metallicities, the ranges shrink dramatically.
 \item Unlike in panel (b), the low-metallicity tracks do not run together at high mass. In other words, high-mass GCs formed from very different metallicity gas might have very similar present-day \feh, but would be distinguishable by different values of $\sigma_0$.
 \item The maximum possible effect of self-enrichment, given by the extent of the tracks, drops strongly at high metallicity. This results in another forbidden region at high \feh\ and high $\sigma_0$ -- large intrinsic dispersions cannot be produced by self-enrichment in very metal-rich clusters.
 \item Neglecting the curvature induced by the ``bump'', the overall slope of the relation ranges from $\sigma_0 / \Delta \feh \approx 0.3$ at $\fehinit=-3$ to $\sigma_0 / \Delta \feh \approx 0.7$ at $\fehinit=-0.5$.
\end{enumerate}

\section{Comparison to Observations}\label{sec:obsdata}

\subsection{Method and Results}\label{sec:obsresults}

\begin{figure}
\plotone{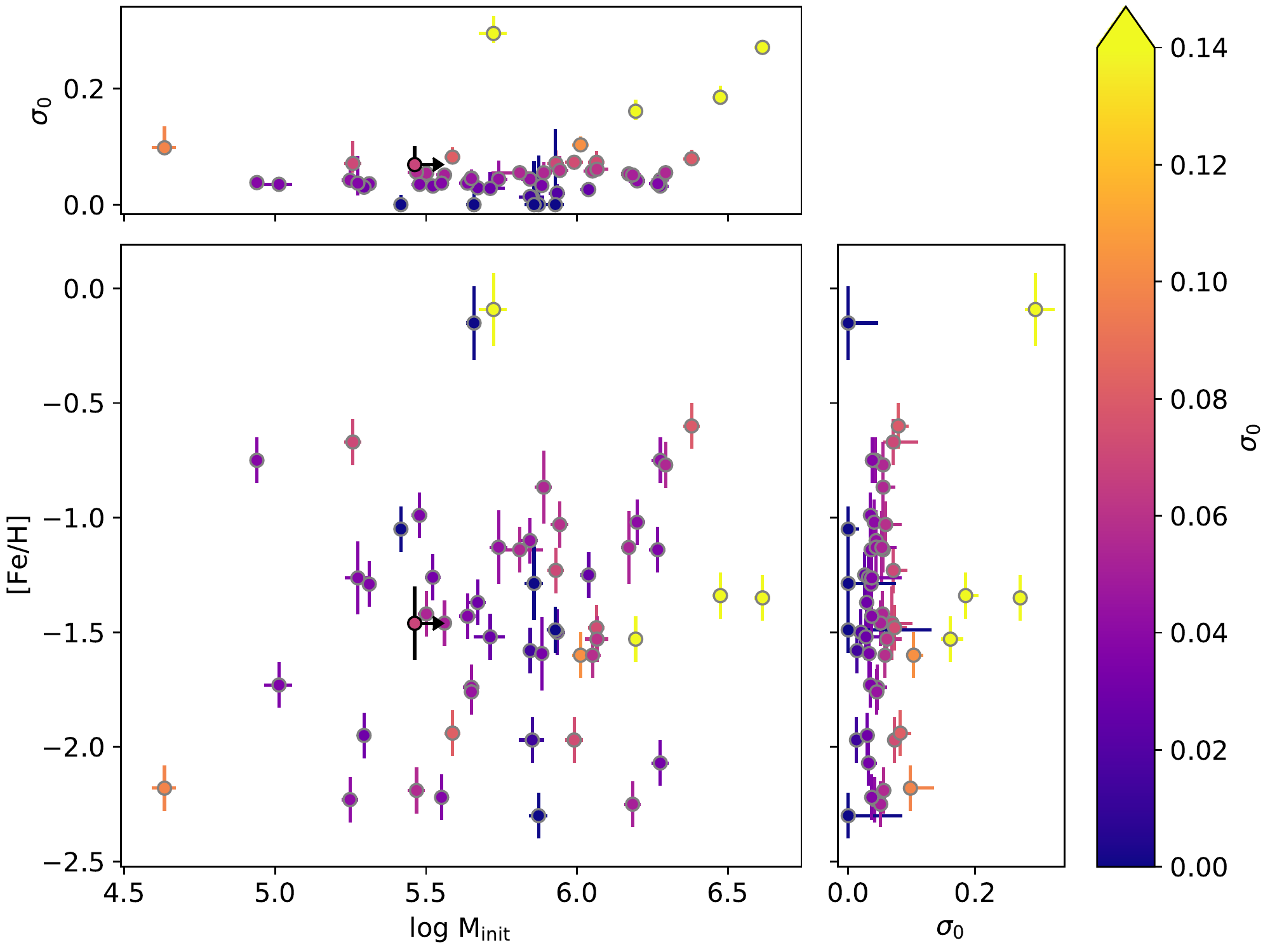}
\caption{\label{fig:obssigma}%
Properties of Milky Way GCs. The main panel shows the current iron abundance \feh, from the catalog of \citet{Kruijssen19-catalog}, as a function of the initial cluster mass \Minit, as derived by \citet{Balbinot18}, colored by the internal iron abundance spread $\sigma_0$ derived by \citetalias{Bailin19} and updated in this paper. The panels on the top and right show $\sigma_0$ as a function of \Minit\ and [Fe/H] respectively. FSR~1758 is shown with a black lower limit symbol on \Minit.%
}
\end{figure}

Comparing models of iron abundance spreads to observations requires a
catalog of homogeneously derived measurements from high spectral resolution data that are careful to distinguish the impact of statistical versus systematic uncertainty in the abundances for individual stars. Until recently, such a catalog was lacking.
\citet{Bailin19} (hereafter \citetalias{Bailin19}) developed a method to measure the internal dispersion $\sigma_0$ from literature high resolution spectroscopic \feh\ measurements of RGB stars while taking into account both random and systematic error. The internal dispersions derived using this method are the $1\sigma$ standard deviations of the best fitting Gaussians, and are directly comparable to the model $\sigma_0$ values discussed in Section~\ref{sec:modelpred}. We have updated the \citetalias{Bailin19} catalog using new results from \citet{Marino19}, who obtained VLT FLAMES-UVES spectra for 18 additional stars in NGC~3201, \citet{Villanova19} and \citet{RomeroColmenares21}, who analyzed 9 and 11 stars respectively for FSR~1758,
and \citet{Meszaros20} (hereafter \citetalias{Meszaros20}), who updated the APOGEE measurements from \citet{Masseron19} with data from APOGEE-South. For all GCs that were in \citet{Masseron19}, we replace those data with that of \citetalias{Meszaros20} because the measurements for each star are identical; the only difference is a slight change in the selection of which stars belong to each cluster. For the remaining GCs, we combined the \citetalias{Meszaros20} data with the data used in \citetalias{Bailin19} as described in that paper.

For \citet{RomeroColmenares21}, we used the abundances obtained using the photometric atmospheric parameters, but confirmed that there are no qualitative differences if the spectroscopic parameters are used instead.

The errors quoted in \citetalias{Meszaros20} are sometimes unrealistically low (e.g. 0.001~dex). \citet{Joensson20} used repeat observations to empirically characterize the uncertainty of abundance determinations in APOGEE based on a star's signal-to-noise ratio, metallicity, and effective temperature. For each star, we replace the quoted uncertainty from \citetalias{Meszaros20} with the value calculated using \citet{Joensson20}'s equation~(6) using the parameters from their Table~8 if it is larger. This only affects a small number of stars.

We only use stars from \citetalias{Meszaros20} with quoted $S/N>200$ due to issues with the \feh\ determinations we found for lower $S/N$ stars (see Appendix~\ref{ap:M20SN}). As a result, NGC~5466 and NGC~7089, which were both in the \citetalias{Bailin19} catalog based on data from \citet{Masseron19}, are excluded here because there are no longer a sufficient number of stars. We also examined each cluster in the \feh-$v_{\mathrm{helio}}$ space to remove obvious interlopers (the criteria in \citetalias{Meszaros20} were intentionally not too restrictive, but the resulting interlopers inflate the measured dispersion). A list of interlopers that were removed is given in Appendix~\ref{ap:M20-interlopers}.

Given the $>800$ stars in the \citet{WillmanStrader12} analysis of NGC~5139 ($\omega$~Cen) and the consequently very well-determined value of $\sigma_0$, we did not redo its analysis. For NGC~1851, where \citetalias{Bailin19} quoted the \citet{WillmanStrader12} value, we went back to the original \citet{Carretta11} data (using the UVES and GIRAFFE observations as separate datasets) to complement the new \citetalias{Meszaros20} data.

Note that high quality spectroscopic abundances for stars in NGC~1261 and NGC~6934 have recently been obtained \citep{Munoz21,Marino21}. Both clusters have photometric evidence of hosting multiple populations with different iron abundances \citep{Milone17}, and these recent spectroscopic studies targeted roughly equal numbers of stars from each photometrically-identified population in order to maximize the chance of determining whether the spectroscopic abundances of the populations are statistically different. The results are that in both clusters, the sequences are offset by $0.1$--$0.2$~dex in iron abundance. Although these data are excellent and inform our understanding of iron abundance spreads, they cannot be used for a quantitative measurement of $\sigma_0$ using the method of \citetalias{Bailin19} because of the wildly non-representative target selection --- the photometrically-selected ``anomalous'' stars, which turn out to be iron-rich, are only $<5\%$ of the cluster stars but represent half of the stars that were observed in these studies. The derived value of $\sigma_0$ would therefore be vastly overestimated using our technique.

The properties of the 54 GCs with measurements of $\sigma_0$ are listed in Table~\ref{table:all} and shown in Figure~\ref{fig:obssigma}, which shows $\sigma_0$ as a function of \Minit\ and \feh. \feh\ values have been taken from \citet{Kruijssen19-catalog} where available in order to maximize the homogeneity of the values. For the remaining clusters, we used the values derived in \citetalias{Bailin19} when available, or from \citet{Harris96} (2010 edition) otherwise. For FSR~1758, we used the \feh\ value obtained by our analysis, and used the current mass estimate by \citet{RomeroColmenares21} as a lower limit; given that this cluster lies within the bulge, it is likely that it experienced a large amount of tidal stripping and so \Minit\ is significantly larger. The main panel shows the bivariate distribution colored by $\sigma_0$, while the side panels compare $\sigma_0$ to each variable individually. As noted by \citetalias{Bailin19}, there is an apparent correlation between \Minit\ and $\sigma_0$ at the high-mass end (all GCs with $\log \Minit > 6$ have measurable iron abundance spreads), but not at low mass. There also appears to be an increase in $\sigma_0$ at both high and low metallicity, although it is difficult to separate these correlations due to the small number of clusters.

\begin{figure}
\plotone{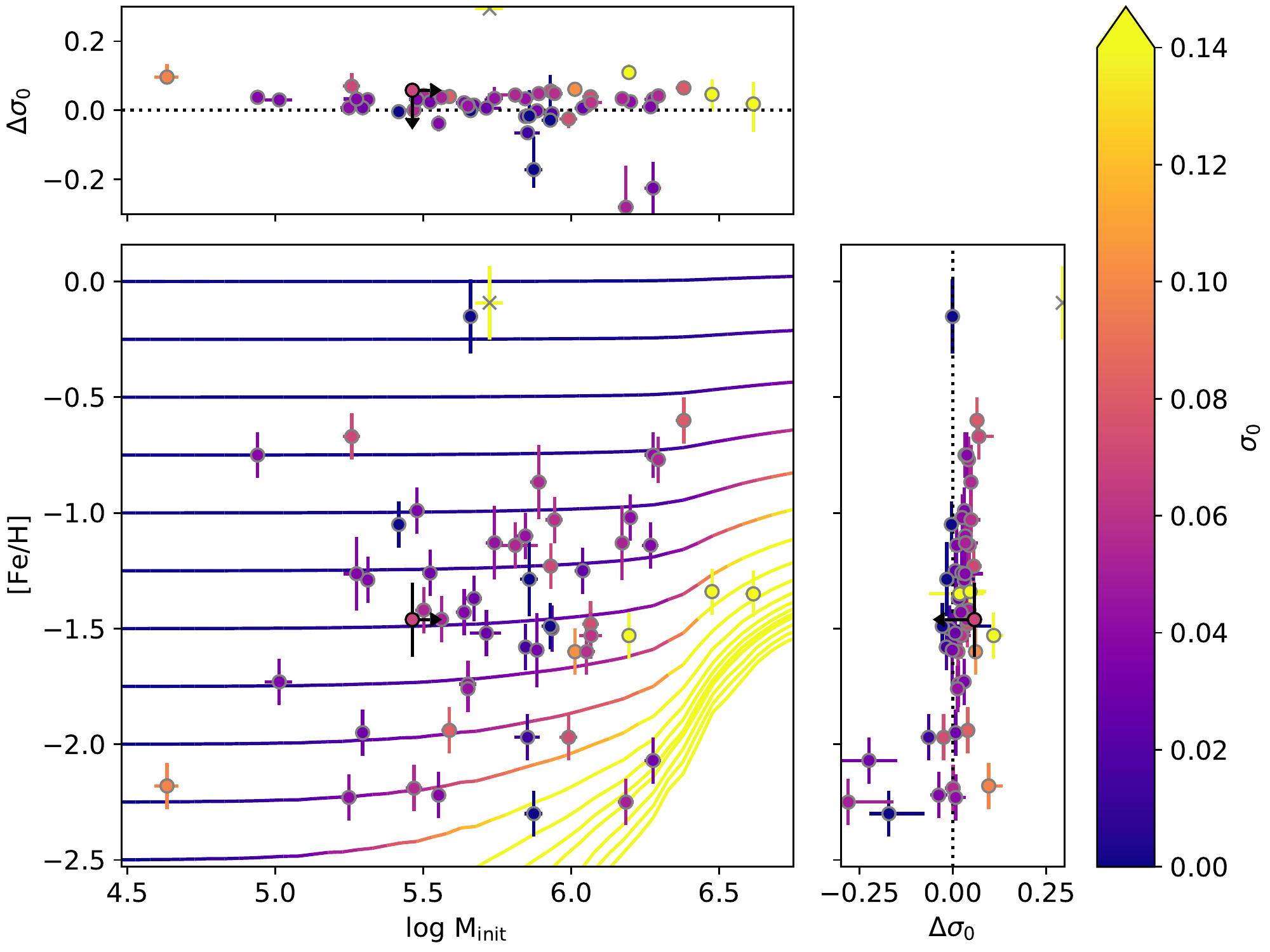}
\caption{\label{fig:predicted sigma comparison}%
Predicted and observed internal iron abundance spreads as a function of initial cluster mass \Minit\ and iron abundance [Fe/H]. The main panel shows the predictions of the \citetalias{Bailin18} clumpy self enrichment model as curves colored by the predicted value of $\sigma_0$, with the observed Milky Way GCs overplotted as in Figure~\ref{fig:obssigma}. The panels on the top and right show the difference between the observed and predicted values of $\sigma_0$ ($\Delta \sigma_0 \equiv \sigma_0^{\mathrm{obs}} - \sigma_0^{\mathrm{pred}}$) as a function of the initial cluster mass and [Fe/H] respectively. FSR~1758 is shown with black upper/lower limit symbols (\Minit\ is a lower limit so $\sigma_0^{\mathrm{pred}}$ is also a lower limit and $\Delta \sigma_0$ is an upper limit).
Terzan~5, which cannot be explained in the GC self enrichment model (Section~\ref{sec:terzan5}), is marked with an x.%
}
\end{figure}

In Figure~\ref{fig:predicted sigma comparison}, we compare those observations to the model. Each track corresponds to a value of \fehinit, with the segments along each track colored by the predicted value of $\sigma_0$. This can be directly compared to the colors of the data points, which designate the observed values as in Figure~\ref{fig:obssigma}. The largest values of $\sigma_0$ are predicted to occur at high \Minit\ and low \feh, as discussed in Section~\ref{sec:modelpred}, and indeed most of the high-$\sigma_0$ clusters do lie towards that direction in parameter space, although there are clearly exceptions (Terzan~5, denoted with an x, is discussed further in Section~\ref{sec:terzan5}; for now, we merely note that it is an extreme outlier).

We predict the value of $\sigma_0$ for each cluster by performing bilinear interpolation of the model tracks in this plane. Uncertainties come from two sources: observational uncertainty in \Minit\ and \feh, and the stochasticity of the model. The effects of observational uncertainty are estimated using 1000 Monte Carlo samples from a two dimensional Gaussian around each point in this parameter space and taking the 16th and 84th percentiles of the resulting $\sigma_0$ distribution. The model stochasticity is shown by the size of the shaded band in Figure~\ref{fig:FeH_model_tracks}c; this is interpolated between model tracks for specific GCs. We found no cross terms between the observational and model uncertainties, so the final uncertainty on $\sigma_0$ is simply their sum in quadrature.

The side panels show the difference between the observed and predicted value $\Delta \sigma_0 \equiv 
\sigma_0^{\mathrm{obs}} - \sigma_0^{\mathrm{pred}}$. There are several points worth noting:

\begin{enumerate}
 \item There is a median offset of $\Delta \sigma_0 = 0.024$, which is larger than the median uncertainty in $\Delta \sigma_0$ of $0.015$. In other words, the model underpredicts the observed iron abundance spread. This is a substantial fraction of the mean observed value of $\sigma_0 = 0.057$, and suggests that some of the internal iron abundance spread comes from sources other than self enrichment, such as inhomogeneities in the protocluster gas cloud; it may also point to deficiencies in the \citetalias{Bailin18} model.
 \item The model captures the mass dependence of $\sigma_0$: $\Delta \sigma_0$ does not show any trend with \Minit.
 \item With the exception of three low-$\sigma_0$ outliers (see below), the model captures the increase in $\sigma_0$ at low metallicity: $\Delta \sigma_0$ does not show a trend with \feh\ there. However, the increase in $\sigma_0$ at the high \feh\ end is not captured by the model. It is possible that this might be alleviated by modifying some of the model parameters, but it is inherently difficult for the model to produce a large internal dispersion at high metallicity. However, the model assumes that none of its parameters depend on metallicity; if, for example, there is a significant metallicity dependence to the core collapse supernova iron yield, that could potentially improve the agreement.
 \item There are three high-mass low-metallicity clusters with spreads that are much lower than predicted by the model: NGC~2419 ($\sigma_0^{\mathrm{pred}} = 0.26^{+0.18}_{-0.08}$, $\sigma_0^{\mathrm{obs}}=0.03\pm 0.01$), NGC~6341 ($\sigma_0^{\mathrm{pred}}=0.17^{+0.04}_{-0.06}$, $\sigma_0^{\mathrm{obs}} < 0.09$), and NGC~7078 ($\sigma_0^{\mathrm{pred}}=0.33^{+0.22}_{-0.11}$, $\sigma_0^{\mathrm{obs}}=0.051^{+0.008}_{-0.002}$). These clusters lie in the ``bump'' in Figure~\ref{fig:FeH_model_tracks}c ($\log \Minit \sim 6$, $\fehinit < -2$) where the predicted value of $\sigma_0$ rises and then falls. As noted in Section~\ref{sec:modeldiscussion}, this feature is quite sensitive to the model details and so the predicted $\sigma_0$ might be overestimated in this regime.
\end{enumerate}

\subsection{Discussion}\label{sec:self-enrichment-discussion}

Although the present work is focused on estimating the effects of core collapse supernovae on iron abundances, rather than providing a comprehensive model of self enrichment in GCs,
it is worth discussing implications that the success of the
model has to understanding the full phenomenon in its richness.

The main conclusion of Section~\ref{sec:obsresults} is that for most GCs, the observed spread in iron abundances within the GC is consistent with being produced by retention of core collapse supernova ejecta within a clumpy star formation region, where the clumps within the star forming cloud have different amounts of ejecta incorporated in them and therefore form stars with different iron abundances. Although we have parameterized this in terms of a single abundance spread, $\sigma_0$, the distinct clumps often lead to distinct subpopulations, as noted by \citetalias{Bailin18}, and observed in many GCs (e.g. \citealp{Bellini17}).

However, it seems unlikely that this same process is responsible for the internal variations of all of the elements. For example, although core collapse supernovae produce copious amounts of oxygen, the observed oxygen abundances are not correlated with the iron abundances in clusters where both elements have wide enough variation to test for a correlation \citep{Marino15}. Another process that occurs on a different timescale must be required.

A more interesting situation is that of the s-process elements. GCs with large iron abundance spreads also often show significant variation in these (the so-called Type~II clusters; \citealp{Marino15,Milone17}), with the iron and s-process abundances correlated between stars within a cluster. In the context of the GCZCSE model, where the variation occurs due to the existence of a clumpy medium, this would be possible if much of the enrichment of s-process elements occurs on a timescale when protocluster retains the same clumpy structure. Interestingly, this requires the s-process enrichment to occur relatively rapidly -- not necessarily in the supernovae themselves, but on timescales not much longer than tens of Myr, which is shorter than many of the sources commonly thought of as
locations of s-process production like AGB stars, and suggests that massive stars
might be a significant contributor.

\section{Interpreting Observed Globular Clusters Using the Self Enrichment Model}\label{sec:fehinit}

\subsection{Method}

In the \citetalias{Bailin18} self enrichment model, the present-day \feh\ and $\sigma_0$ are a consequence of the original cluster properties \fehinit\ and \Minit. We can therefore use the observed properties of GCs to reconstruct their properties as protoclusters.

\begin{figure*}
\plotone{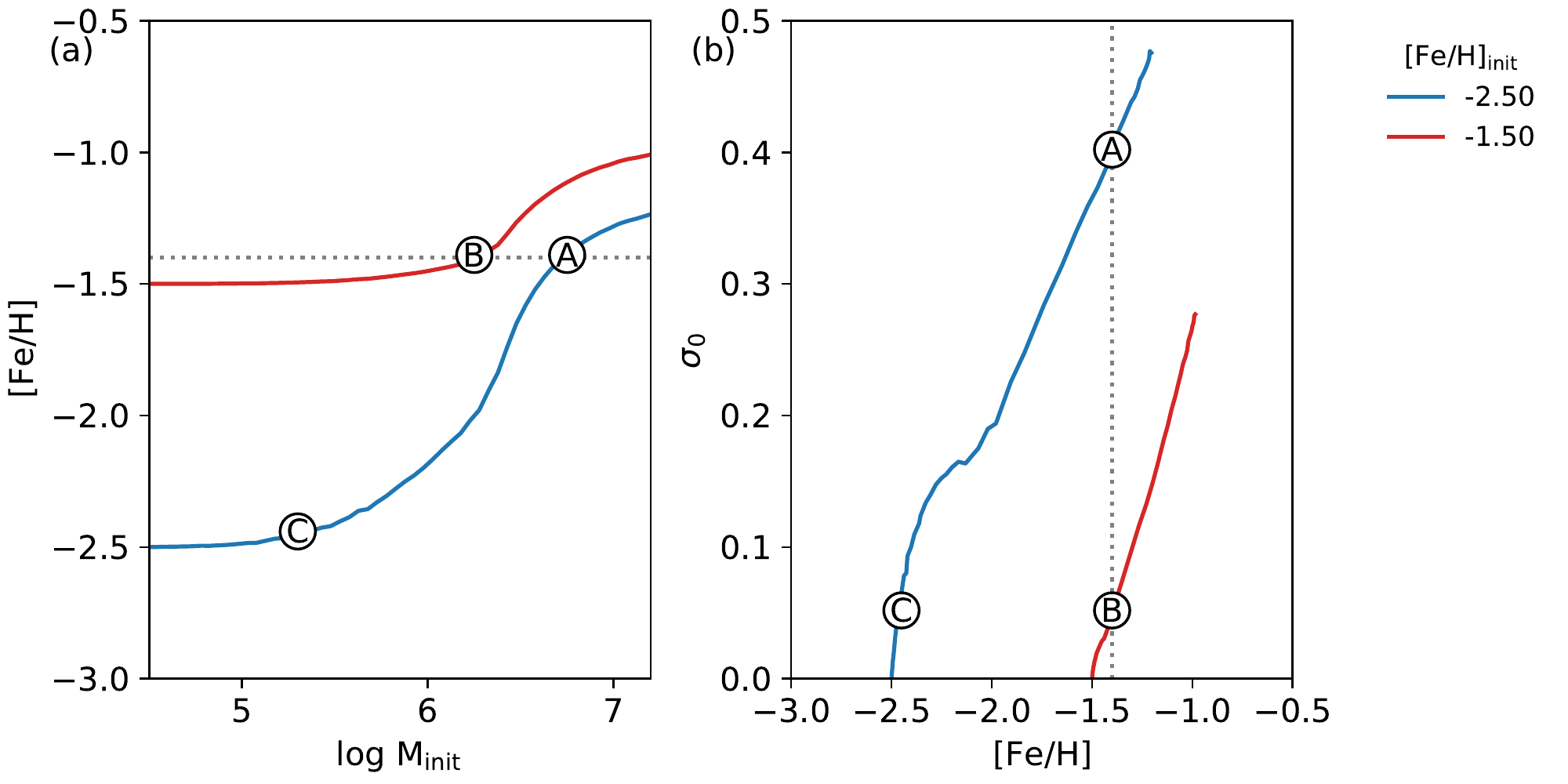}
\caption{\label{fig:reconstruct schematic}%
Schematic view of how to reconstruct \fehinit\ of a GC given its present-day observables \feh, $\sigma_0$, and \Minit. The clusters labeled A and B both have $\feh=-1.4$ (gray dotted line) while C has $\feh=-2.45$, leading one to naively suspect that A and B were born in similar circumstances that were very different from where C was born. However, A's high internal spread and mass indicate that it has experienced a very large amount of self enrichment; in fact, both clusters A and C came from metal-poor gas that had $\fehinit=-2.5$ (blue line), while cluster B formed from more metal-rich gas that had $\fehinit=-1.5$ (red line). The true value of \fehinit\ can be reconstructed for a cluster by finding which model track it lies on.%
}
\end{figure*}

This is demonstrated in the schematic Figure~\ref{fig:reconstruct schematic}, showing three example GCs. Two of them, labeled A and C, formed from identical low-metallicity gas, $\fehinit=-2.5$, and the other, B, formed from much more metal-rich gas, $\fehinit=-1.5$. However, one of the metal-poor gas clouds, A, formed a very massive GC with a large amount of self enrichment, reflected in a much larger internal dispersion $\sigma_0$; in fact, it self-enriched so much that its present-day metallicity is as high as, B, which formed from metal-rich gas. If we looked purely at the present-day \feh, we might naively conclude that the metal-poor cluster C formed early from metal-poor gas while A and B formed later from metal-rich gas, but the dramatically different values of $\sigma_0$ and \Minit\ tell us a different story. This example shows how we can use $\sigma_0$ to reconstruct the initial metallicity of the gas from which a GC formed, and how it can lead us to quite different conclusions about its history.

In practice, for a given GC with an observed value of \feh\ and $\sigma_0$, we perform bilinear interpolation between the tracks of Figure~\ref{fig:FeH_model_tracks}d to determine the model prediction $\fehinit(\sigma_0)$. Alternatively, for a given GC with an observed value of \feh\ and \Minit, we perform bilinear interpolation between the tracks of Figure~\ref{fig:FeH_model_tracks}b to determine the model prediction $\fehinit(\Minit)$. Most clusters have measurements of \Minit\ from \citet{Balbinot18} and therefore have a determination of $\fehinit(\Minit)$; a smaller subset have measurements of $\sigma_0$ from \citetalias{Bailin19} and Section~\ref{sec:obsdata} and therefore have a determination of $\fehinit(\sigma_0)$. As discussed in \citetalias{Bailin18}, the tracks in Figure~\ref{fig:FeH_model_tracks}d are much more robust to the parameters of the GCZCSE model and are therefore we prefer $\fehinit(\sigma_0)$ when available.

Uncertainties in the determination of \fehinit\ come from two sources: (1) observational uncertainties and (2) model stochasticity.
\begin{enumerate}
 \item Observational uncertainties are determined by Monte Carlo sampling 1000 points in a two dimensional Gaussian around each observed GC in the relevant \feh-$\sigma_0$ or \feh-\Minit\ parameter space and finding the 16th- and 84th-percentile of the resulting \fehinit\ distribution. For cases with asymmetric input error bars, points are sampled from an asymmetric Gaussian distribution with different width above and below the central value.
 \item The effects of model stochasticity are estimated using the tracks that define the boundaries of the shaded regions in Figure~\ref{fig:FeH_model_tracks}. We performed the same bilinear interpolation as above but using the upper and lower boundary tracks instead of the median relation, and used these values as the estimate of the error due to model stochasticity.
\end{enumerate}
The total error estimate for \fehinit\ is the observational and stochastic error added in quadrature\footnote{We neglect cross terms because we found no correlation between the model stochasticity errors and the location of a point within the Monte Carlo cloud.}. This is calculated separately for the upper and lower bounds.
See \citet{recongc} for code that performs this reconstruction.

\subsection{Results}

\begin{deluxetable*}{ccccccccc}
\tablecaption{\label{table:all}Observed and predicted properties of the GC sample.}

\tablehead{\colhead{Cluster Name} & \colhead{Alternate Name} & \colhead{$\log \Minit$} & \colhead{[Fe/H]} & \colhead{$\sigma_0$} & \colhead{$\sigma_0^{\mathrm{pred}}$} & \colhead{$\fehinit(\sigma_0)$} & \colhead{$\fehinit(\Minit)$} & \colhead{Progenitor}}

\startdata
NGC 104 & 47 Tuc & $6.28 \pm 0.03$ & $-0.75 \pm 0.1$ & $0.043^{+0.002}_{-0.003}$ & $0.011^{+0.005}_{-0.003}$ & $-0.82^{+0.09}_{-0.11}$ & $-0.77^{+0.10}_{-0.11}$ & Main \\
NGC 1261 & \nodata & $5.64 \pm 0.03$ & $-1.23 \pm 0.1$ & \nodata & $0.009 \pm 0.003$ & \nodata & $-1.24 \pm 0.10$ & GE \\
NGC 1851 & \nodata & $5.84 \pm 0.03$ & $-1.1 \pm 0.1$ & $0.044^{+0.006}_{-0.001}$ & $0.011 \pm 0.003$ & $-1.18 \pm 0.11$ & $-1.11^{+0.11}_{-0.10}$ & GE \\
NGC 1904 & M 79 & $5.67 \pm 0.03$ & $-1.37 \pm 0.1$ & $0.029^{+0.006}_{-0.003}$ & $0.014 \pm 0.004$ & $-1.43^{+0.10}_{-0.09}$ & $-1.39^{+0.11}_{-0.10}$ & GE \\
\enddata

\tablecomments{Table 1 is published in its entirety in the machine-readable format. A portion is shown here for guidance regarding its form and content.}

\end{deluxetable*}

All information about each Milky Way GC is given in Table~\ref{table:all}. Columns 1 and 2 list the name and alternate names for the cluster, Column 3 lists the initial mass from \citet{Balbinot18}, Column 4 lists our adopted \feh, Column 5 gives the internal iron spread $\sigma_0$, and Column 6 shows the predicted value of $\sigma_0$ from the GCZCSE model. The reconstructed values of \fehinit\ using $\sigma_0$ and \Minit\ are in Columns 7 and 8 respectively, and the proposed progenitor accretion event in the literature (see Section~\ref{sec:age-metal}) is in Column 9.

\begin{figure*}
\plotone{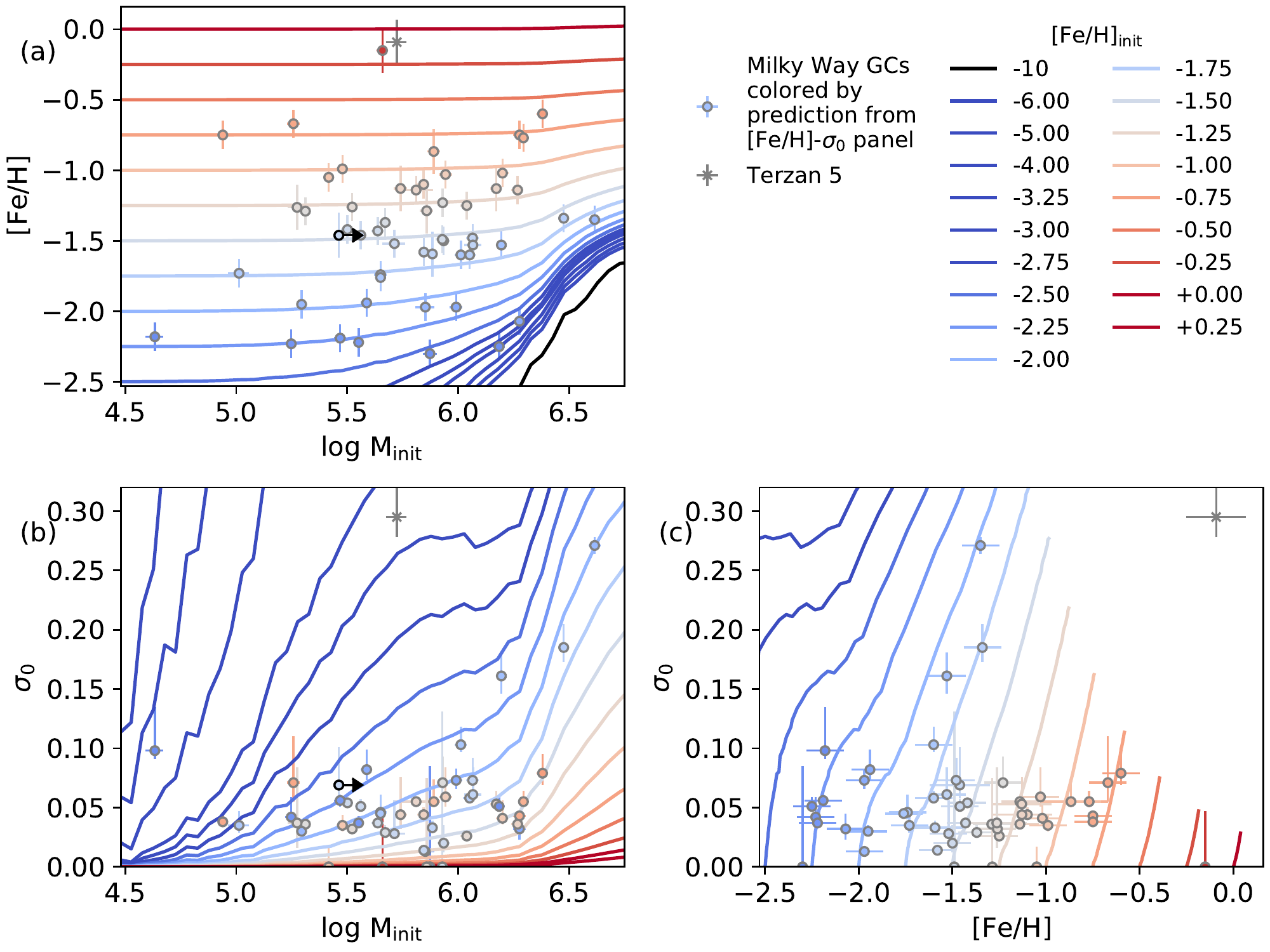}
\caption{\label{fig:sigfehi predictions}%
Distribution of Milky Way GCs in the \Minit-\feh-$\sigma_0$ parameter space. Curves show the predictions of the self enrichment model as in Figure~\ref{fig:FeH_model_tracks}; the uncertainties due to model stochasticity are not shown but are identical to that figure. Data points are colored by $\fehinit(\sigma_0)$, i.e., the value of \fehinit\ interpolated from the model tracks in panel (c) based on the measured internal iron spread $\sigma_0$. The x indicates Terzan~5, which cannot be explained by the self enrichment model.%
}
\end{figure*}

Figure~\ref{fig:sigfehi predictions} shows all of the GCs with measured values of $\sigma_0$ in the \feh-\Minit-$\sigma_0$ parameter space. The curves show the predictions of the \citetalias{Bailin18} self enrichment model (the uncertainties are omitted but are identical to the shaded regions in Figure~\ref{fig:FeH_model_tracks}). Data points are colored by $\fehinit(\sigma_0)$, i.e. by interpolation between tracks in panel (c). Note that aside from Terzan~5 (see Section~\ref{sec:terzan5}), no other GCs lie in the high-\feh\ high-$\sigma_0$ forbidden region at the top-right of panel (c); their internal iron abundance dispersion is explainable by self enrichment during the formation of a GC and does not inherently require a deeper potential well like a dwarf galaxy (see, however, Section~\ref{sec:age-metal}). The colors of the data points also match the model tracks well in panel (a) and, to a lesser degree, panel (b), suggesting that the observed GCs lie near the manifold in the three-dimensional parameter space that is traced by the model tracks.

It is interesting that although the GCZCSE model predicts that it is easiest for clusters with the lowest metallicities to achieve high internal dispersion, the highest-$\sigma_0$ clusters are not those with the lowest metallicity, but have $\feh \sim -1.5$. This is because these clusters have among the highest mass of all of the clusters, as seen in panel (b); the combination of moderately low metallicity and very high mass combine to give a larger dispersion than the low mass lower-metallicity clusters, and the colors of those data points are similar to the colors of the surrounding model tracks.

The bulge cluster FSR~1758 has a lower limit $\log \Minit > 5.46$. We can use the self enrichment model to back out the value of \Minit\ that would be consistent with its observed value of $\sigma_0=0.069$. By interpolating between model tracks, the best fit has $\log \Minit = 6.3$, suggesting that its current mass is a factor of $\sim 7$ times less than its initial mass, i.e. it has lost $\approx 85\%$ of its mass. Although large, this is not an unusual amount of mass loss for a cluster orbiting in the dense inner galaxy \citep{Balbinot18}.

\begin{figure*}
\plotone{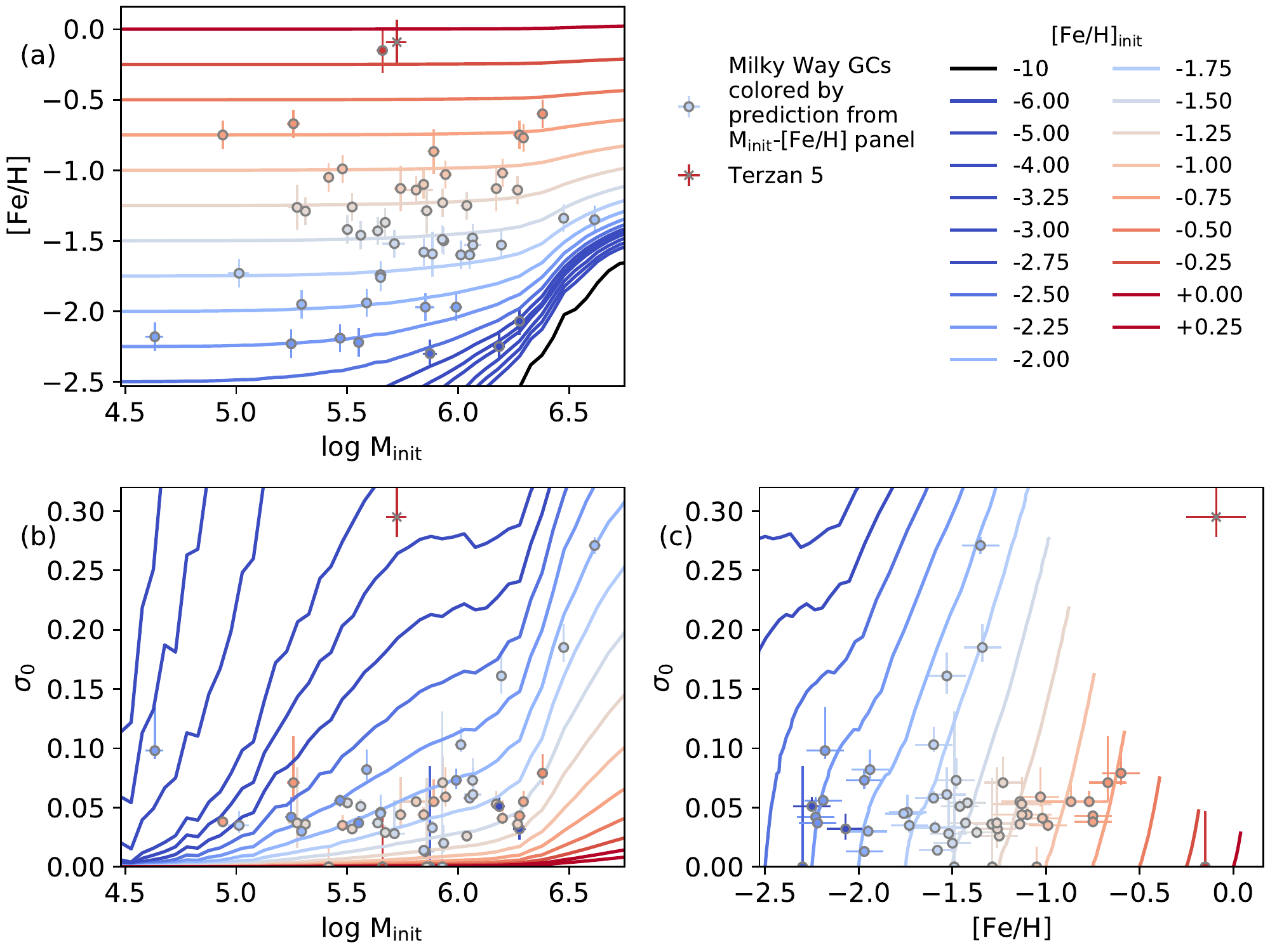}
\caption{\label{fig:mifehi predictions}%
As in Figure~\ref{fig:sigfehi predictions}, but data points are colored by $\fehinit(\Minit)$, i.e. the value of \fehinit\ interpolated from the model tracks in panel (a) based on the mass \Minit. FSR~1758 is not plotted due to the large uncertainty in its \Minit.
}
\end{figure*}

\begin{figure*}
\plotone{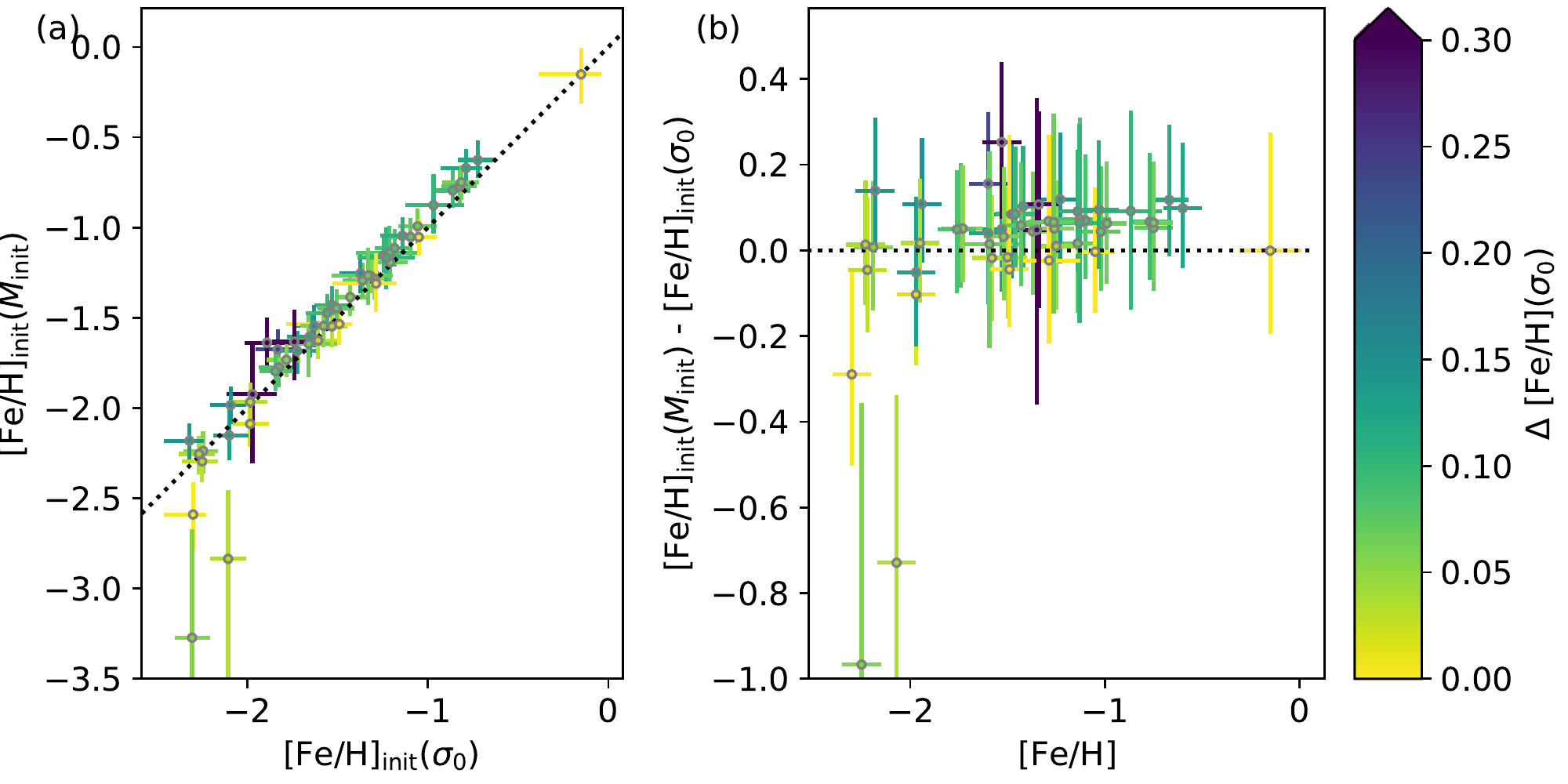}
\caption{\label{fig:misig comparison}%
(a) Comparison between the values of \fehinit\ for Milky Way GCs derived using $\sigma_0$, as in Figure~\ref{fig:sigfehi predictions}, and from \Minit, as in Figure~\ref{fig:mifehi predictions}. Points are colored by the magnitude of the self enrichment correction for the $\sigma_0$ prediction, i.e. $\Delta \fehinit(\sigma_0) \equiv \feh - \fehinit(\sigma_0)$. (b) The difference between the \fehinit\ predictions as a function of cluster metallicity, colored as in panel (a). The predictions using the two methods agree to well within the error bars for most clusters regardless of the magnitude of the self enrichment correction, with the exception of two very low-metallicity clusters that have large uncertainties. There is a systematic shift whereby the predictions using \Minit\ are $0.047 \pm 0.026$~dex higher, but with no strong dependence on \feh\ or $\Delta \feh$.%
}
\end{figure*}

Figure~\ref{fig:mifehi predictions} shows the same information but with data points colored by $\fehinit(\Minit)$, i.e. by interpolation between tracks in panel (a). Note that no GCs lie in the low-\feh\ high-\Minit\ regime that is forbidden by the self enrichment model; no GC has less iron in it than the minimum that would be provided by pure self enrichment in a pristine protocluster cloud. The colors of the data points again match the model tracks well in panel (c) and, to a lesser degree, panel (b): the two methods of determining \fehinit\ give consistent answers. This is explored further in Figure~\ref{fig:misig comparison}, which directly compares them. Panel (a) demonstrates the correlation explicitly. The only two GCs that appear to fall off the relation are two low-\feh\ clusters that are in the regime where the model tracks bunch up and consequently the uncertainty in $\fehinit(\Minit)$ is unusually large; these are also the two largest outliers in Figure~\ref{fig:predicted sigma comparison}. The data points tend to lie slightly above the one-to-one line; this can be seen more clearly in panel (b) which plots the offset between the two determinations as a function of metallicity. The $\fehinit(\Minit)$ values are systematically offset 0.049~dex higher. This is independent of the magnitude of the model correction: the data points are colored by $\feh - \fehinit(\sigma_0)$. This is another manifestation of the fact that the model systematically underpredicts the iron abundance spread (Figure~\ref{fig:predicted sigma comparison}). Note that $\fehinit(\sigma_0)$ and $\fehinit(\Minit)$ are not truly independent since both depend on \feh, and so the scatter is smaller than the errorbar on each individual determination.

\begin{figure}
\plotone{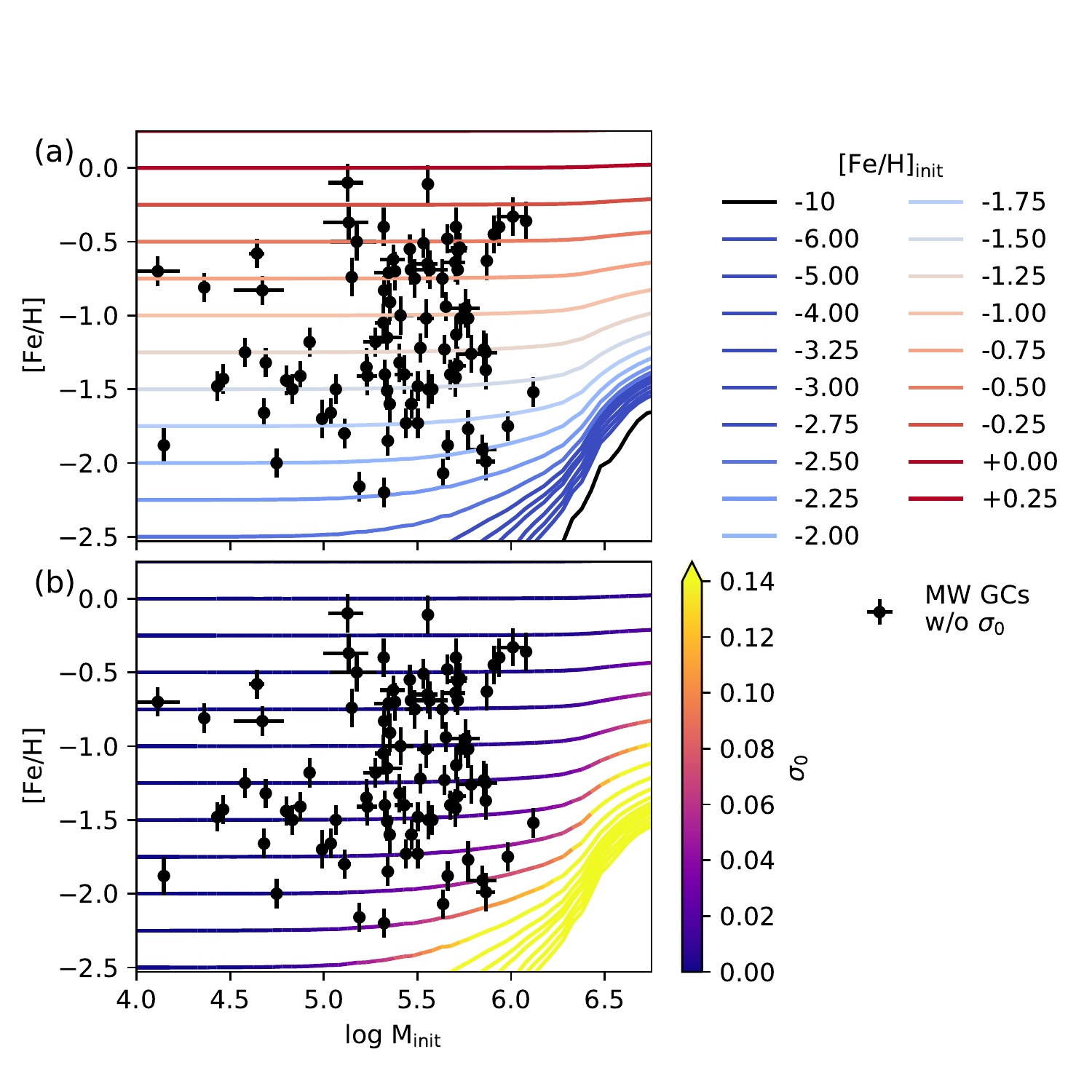}
\caption{\label{fig:unobserved}%
Distribution of Milky Way GCs for which there is no measured value of $\sigma_0$ in initial cluster mass \Minit\ vs. [Fe/H]. Panel (a) shows the self enrichment model tracks for different values of \fehinit, as in Figure~\ref{fig:sigfehi predictions}, while panel (b) shows the model tracks colored by the predicted value of $\sigma_0$, as in Figure~\ref{fig:predicted sigma comparison}. The predicted values for \fehinit\ and $\sigma_0$ for these clusters in Table~\ref{table:all} are taken by interpolating between the model tracks in these plots.%
}
\end{figure}

The 92 GCs with measured \feh\ and \Minit\ but no measured value of $\sigma_0$ are shown in Figure~\ref{fig:unobserved}, along with model track predictions of $\fehinit(\Minit)$ and $\sigma_0$. These clusters tend to be less massive and more metal-rich than for those clusters with measured $\sigma_0$, which is entirely due to selection effects -- detailed distributions of chemical abundances are harder to obtain towards the bulge, which contains more metal-rich GCs, and it is difficult to obtain a good sample of luminous RGB stars for lower-mass clusters. Still, it is notable that even including these GCs does not reveal any in the forbidden bottom-right corner of the \feh-\Minit\ parameter space. Predicted values of $\fehinit(\Minit)$ and $\sigma_0^{\mathrm{pred}}$ for these clusters are again taken by bilinear interpolation between the model tracks, with uncertainties determined as above.

Given that the predictions of the GCZCSE model in \feh-$\sigma_0$ space are much more robust to the model parameters \citepalias{Bailin18}, we use $\fehinit(\sigma_0)$ as the best determination of \fehinit\ when it is available, and use $\fehinit(\Minit)$ otherwise.

The values $\sigma_0^{\mathrm{pred}}$ can be used to guide new observational campaigns looking for multiple populations and iron abundance variations in GCs. There are no missing clusters with very large predicted iron spreads --- all clusters without measured values of $\sigma_0$ have $\sigma_0^{\mathrm{pred}} < 0.1$, which is perhaps not surprising given that observations have already targeted most of the most massive clusters. However, there are clusters in the bottom-right region of Figure~\ref{fig:unobserved}b that have predicted spreads that could be measurable: NGC~6293 has $\sigma_0^{\mathrm{pred}} = 0.084$, and NGC~6287, NGC~6541, and NGC~6779 all have $\sigma_0^{\mathrm{pred}} \sim 0.06$. We suggest these as interesting targets for future observations.

\subsection{Terzan 5}\label{sec:terzan5}

Terzan~5 is the one GC that unambiguously cannot be explained by the GCZCSE model. It has long been known as a wildly discrepant system with subpopulations that differ not only in iron abundance but also by approximately 7~Gyr in age, and with abundance patterns that look more like bulge field stars than the multiple populations of GCs \citep{Origlia11,Origlia19,Massari14,Ferraro16}.

Terzan~5's location deep in the bulge suggests that its formation was dominated by its environment rather than forming as a relatively isolated system as envisioned in the GCZCSE model. Among the possibilities that have been considered in the literature are that it was simply a particularly dense piece of the proto bulge, that it is the remnant of a stripped dwarf galaxy, or that it is the product of a merger between two GCs or a GC and a giant molecular cloud \citep{Massari14,McKenzie18,Pfeffer21}.
Regardless, it is clear that its formation was significantly different enough from that of the other GCs that the GCZCSE model does not provide a good lens for understanding it, and so we neglect it for the remainder of this paper.

\section{Age-Metallicity Relation of Milky Way Constituents}\label{sec:age-metal}

\begin{figure*}
\plotone{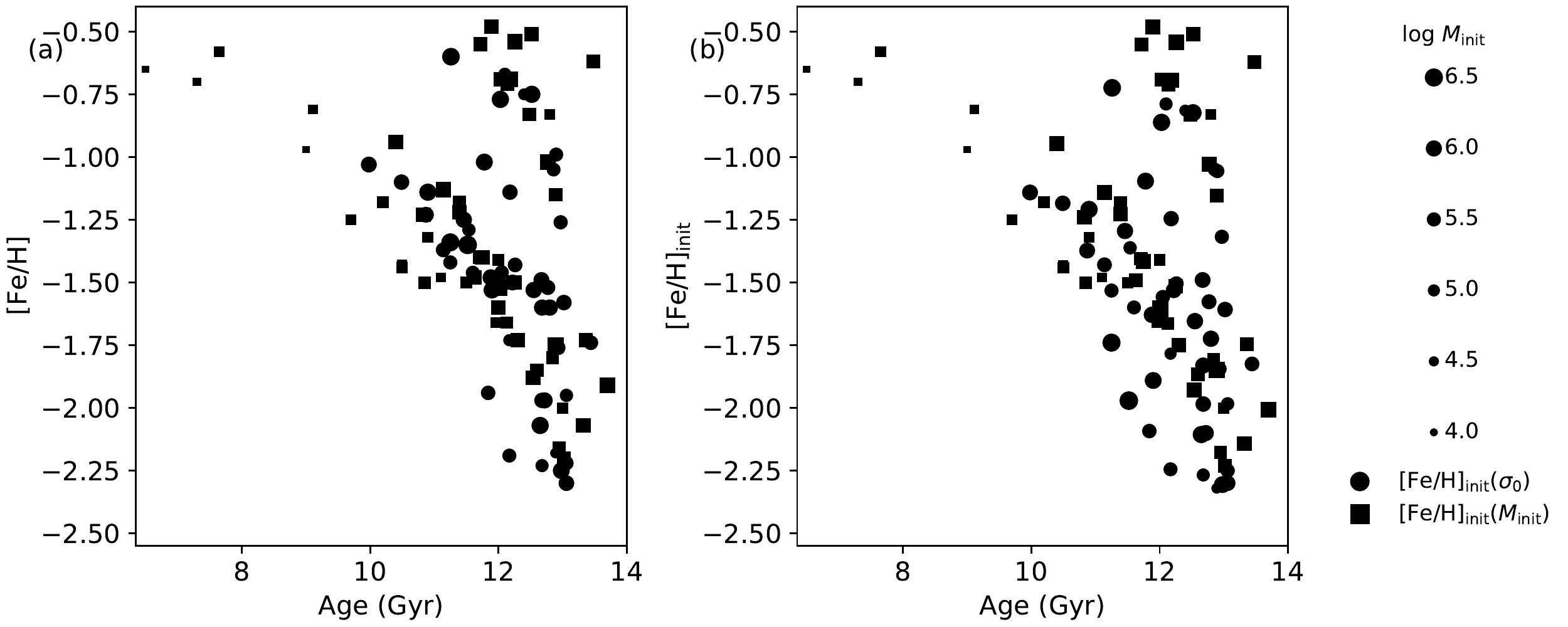}
\caption{\label{fig:agemetal all}%
(a) Relationship between the age and metallicity of Milky Way GCs. Data are taken from \citet{Kruijssen19-catalog}. Symbol sizes denote the initial cluster mass \Minit\ derived by \citet{Balbinot18}. (b) As in the left panel, but using the initial metallicity \fehinit\ as predicted in this paper by the clumpy self-enrichment model. Circles denote clusters with measured values of $\sigma_0$ that were used to predict \fehinit, while squares denote clusters where the less robust \Minit-based method was used.%
}
\end{figure*}

The age-metallicity relation (AMR) of all clusters for which we have determined \fehinit\ are shown in Figure~\ref{fig:agemetal all}. The left panel shows the observed \feh\ values, with the metal-rich branch at the top and metal-poor branch at the bottom readily apparent. The right panel shows the initial metallicities, \fehinit. The overall features from the left panel are strongly preserved in the right panel, which is not surprising since the self-enrichment correction is small for most clusters (67\%\ have corrections of less than 0.02~dex). However, there are clearly differences -- of the remaining clusters, the median correction is 0.08~dex, and they range up to 0.62~dex for NGC~5139 ($\omega$~Cen). The overall consistency of these two panels means that the framework of the Milky Way's accretion history that has been built up by studying the AMR of GCs to date \citep[e.g.][]{Kruijssen20} is robust to the impact of self enrichment.

GCs have been assigned to progenitors based on Gaia DR2 kinematics by \citet{Massari19}, with a few modifications as per \citet{Kruijssen20,Yuan20,Naidu20}:
\begin{itemize}
 \item The ``low energy'' clusters have been relabeled ``Kraken'' \citep{Kruijssen20}.
 \item The Main-Disk and Main-Bulge populations are analyzed together and labeled ``Main''.
 \item Pal 1 is considered ambiguous Main/H-E.
 \item NGC~6441 is considered ambiguous Kraken/Main.
 \item E~3 is considered ambiguous H99/Main.
 \item NGC~6121 is considered ambiguous Kraken/Main.
 \item NGC~5024 and NGC~5053 are considered ambiguous Wukong/H99. 
\end{itemize}

The ``high energy'' clusters are thought to belong to the ensemble of low mass accretion events that each only brought in one or two clusters, and therefore are not expected to show a unique chemical enrichment track \citep{Massari19,Kruijssen20}.

\begin{figure*}
\plotone{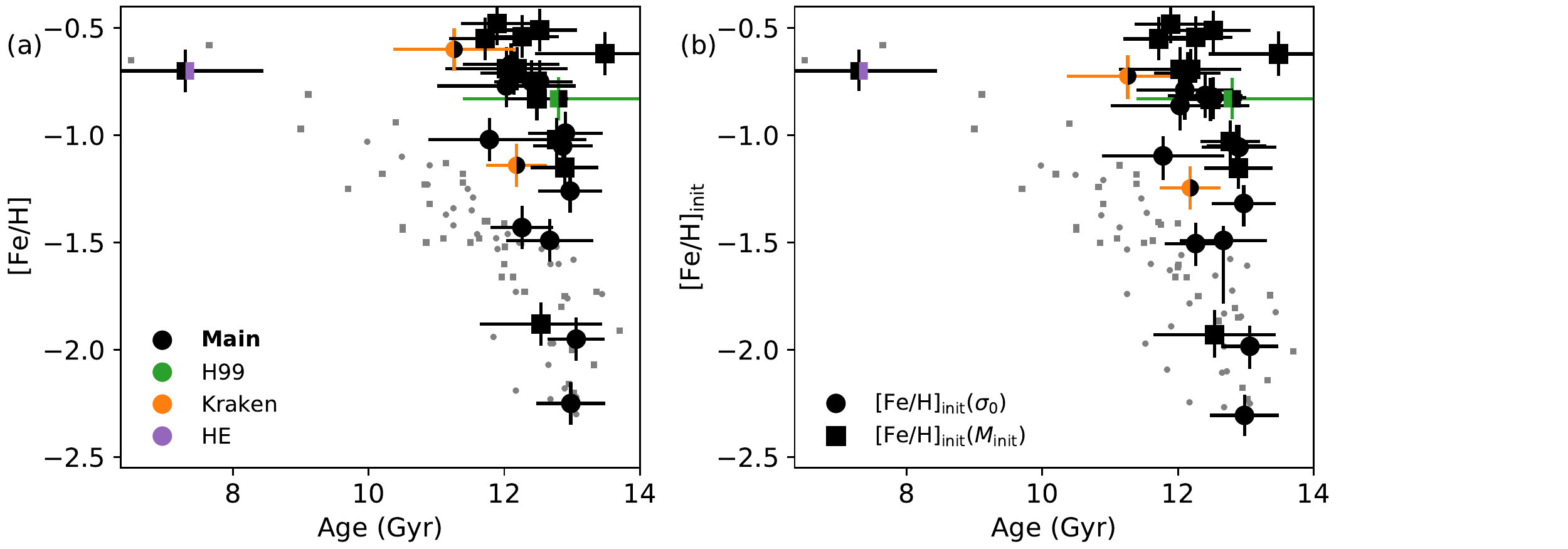}
\plotone{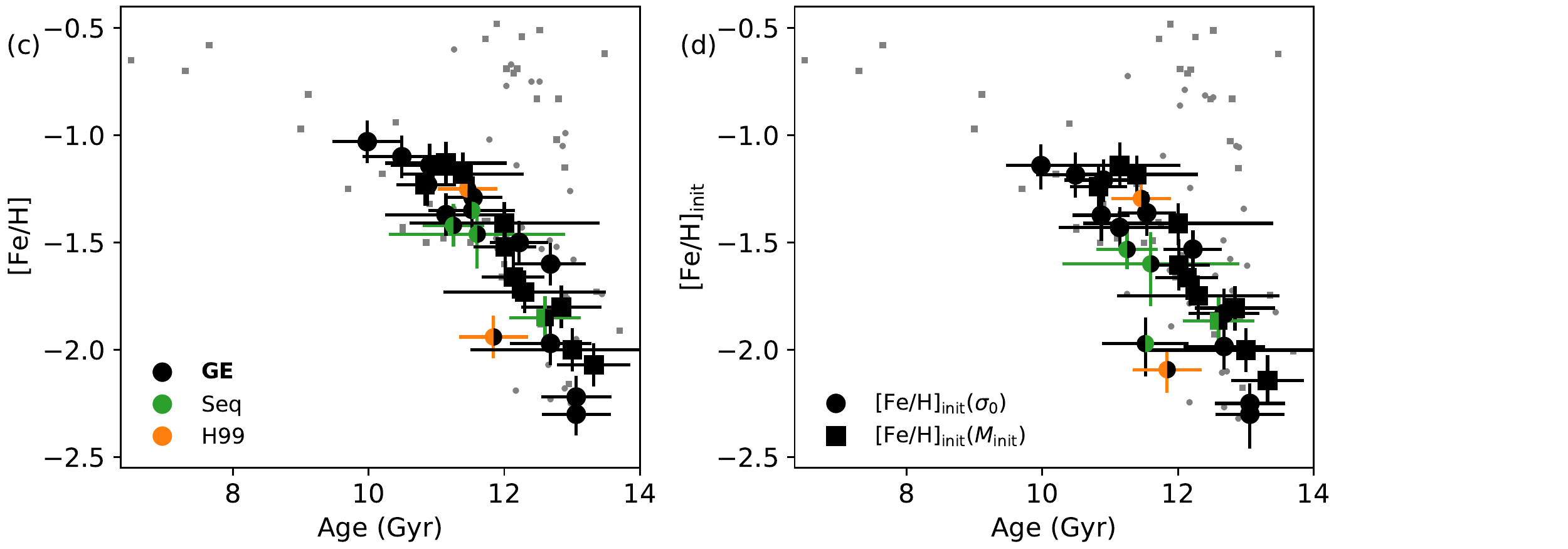}
\plotone{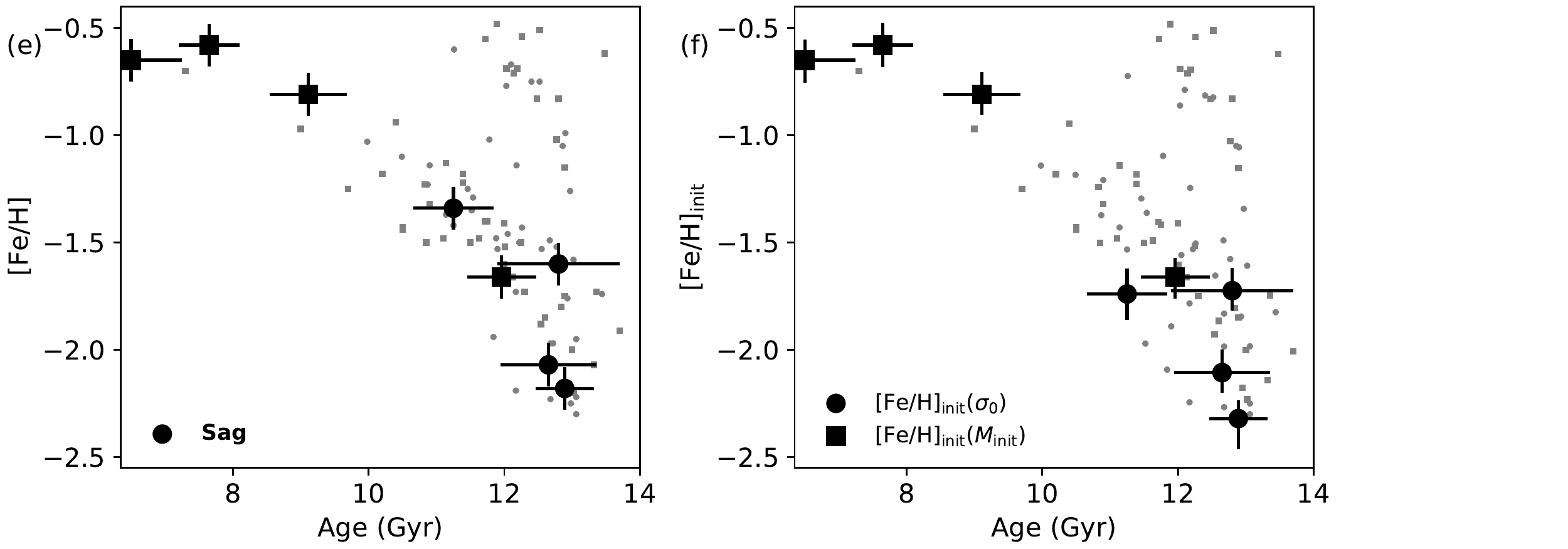}
\plotone{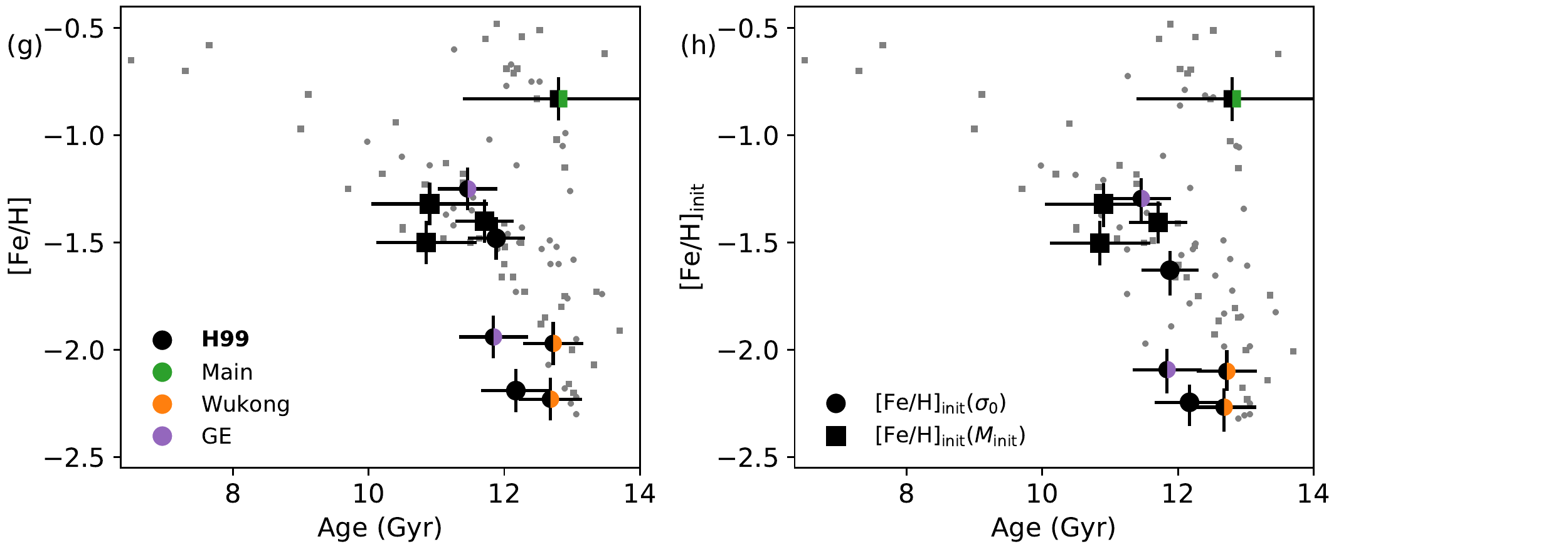} 
\caption{\label{fig:agemetal progenitors}%
Panels (a) and (b) show the age-metallicity relationship of GCs associated with the main progenitor of the Milky Way in large black symbols, with small gray points denoting clusters from other progenitors. GCs with ambiguous assignments are denoted by multicolored points. The left panel plots each cluster at its current [Fe/H], while the right panel show the predicted \fehinit, as in Figure~\ref{fig:agemetal all}. The remaining panels show the clusters associated with other progenitors; progenitor assignment mostly follows \citet{Massari19} except as noted in the text.%
}
\end{figure*}

\addtocounter{figure}{-1}

\begin{figure*}
\plotone{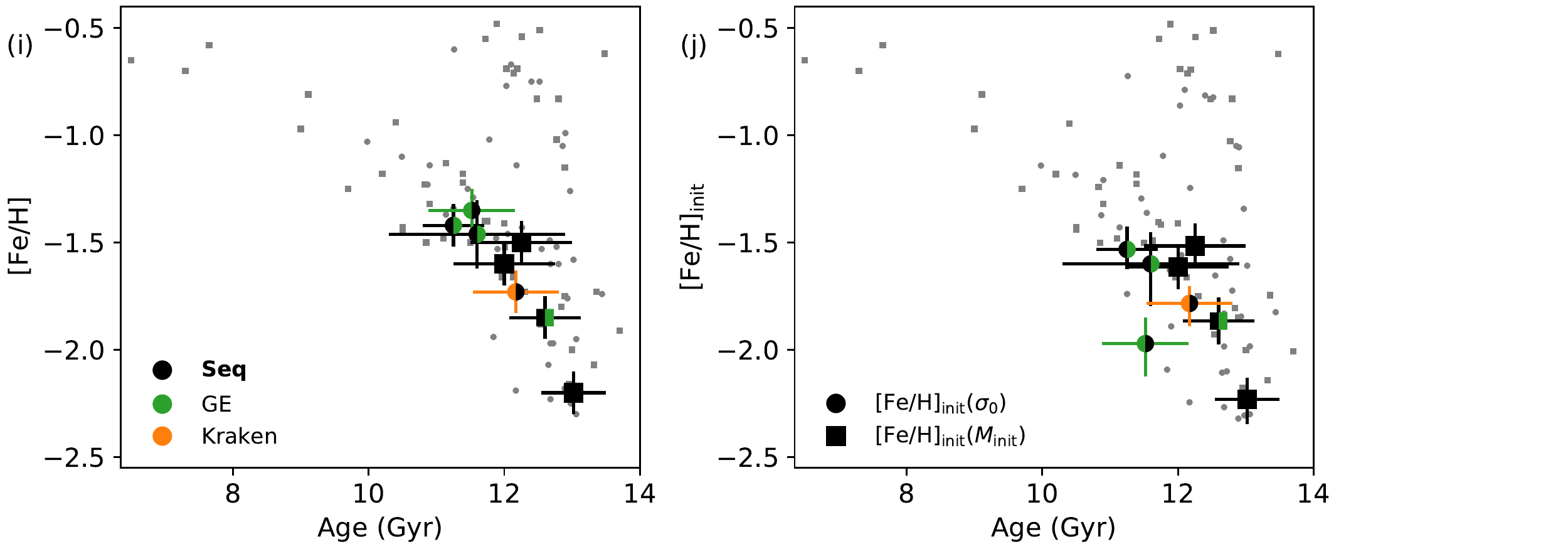}
\plotone{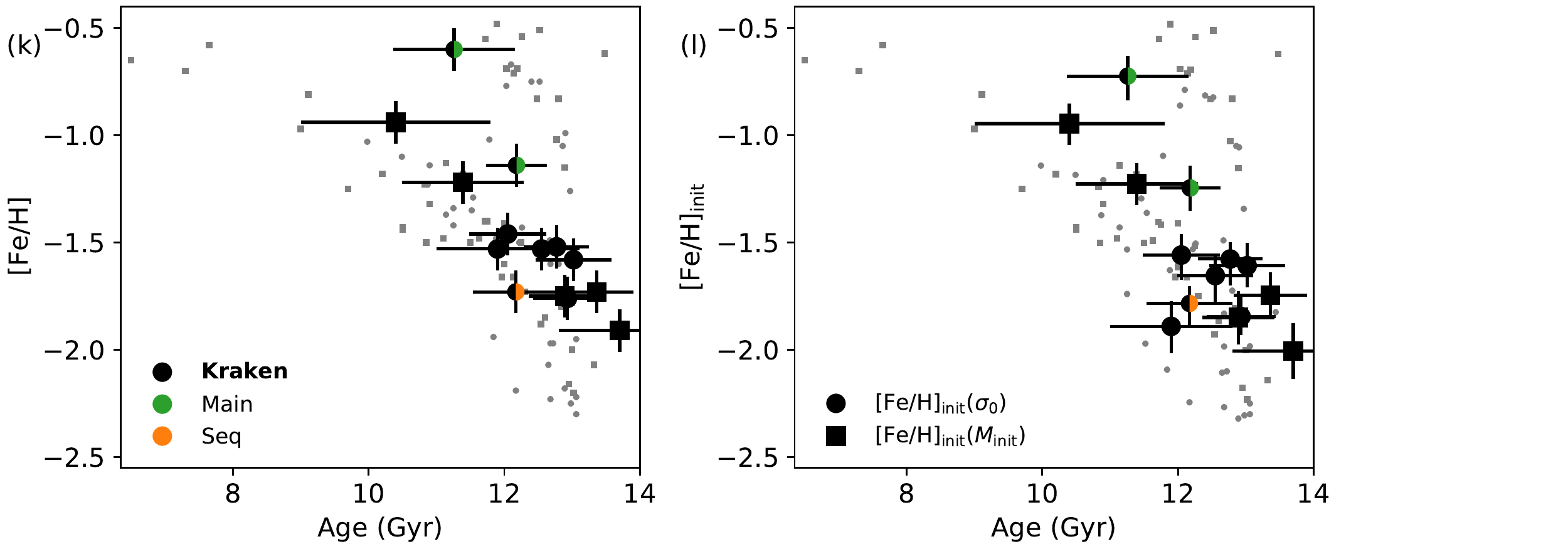}
\plotone{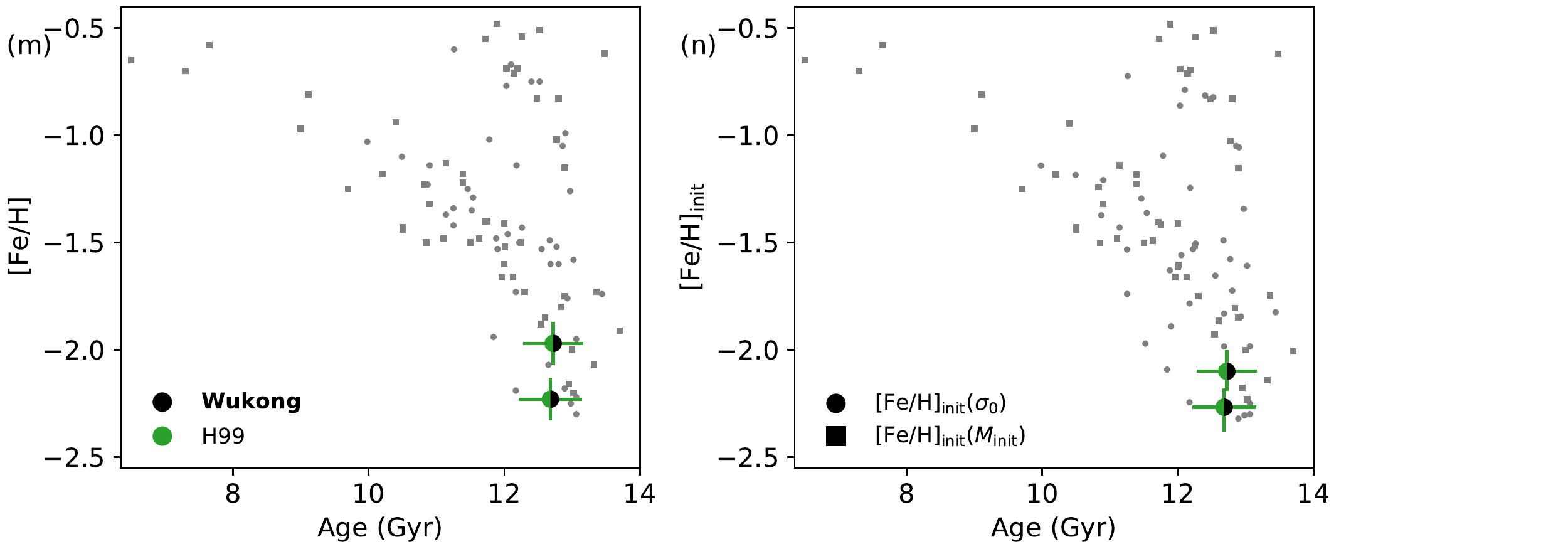}
\plotone{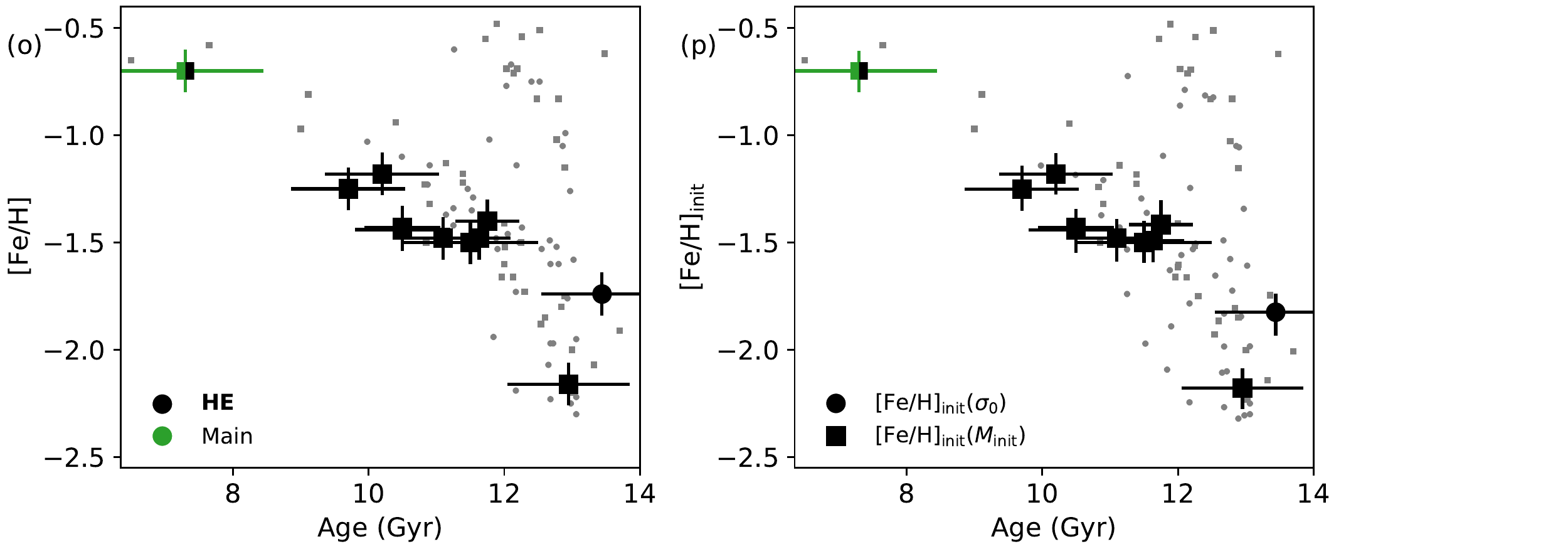} 
\caption{Continued.}
\end{figure*}

The AMR of the GCs assigned to each progenitor is shown in Figure~\ref{fig:agemetal progenitors}. Each panel shows the background distribution of all Milky Way GCs as small gray points. GCs with ambiguous assignments are shown as multicolored points. Once again, the overall features of each AMR are preserved whether looking at \fehinit\ or \feh, although individual clusters do move.

It is interesting to note that for the larger progenitors Gaia-Enceladus and Kraken, what appears to be a single age-metallicity sequence with some spread when looking at \feh\ appears more like distinct parallel sequences when using \fehinit. The Helmi streams also potentially show two sequences, apparent in both \feh\ and \fehinit. This substructure within each progenitor's AMR could be an indication of their own complex history before being accreted onto the Milky Way.

\citet{Pfeffer21} identified NGC~5139 ($\omega$~Cen), NGC~6273, and NGC~6715 (M54) as nuclear star clusters (NSCs) of the galaxies associated with the Gaia-Enceladus, Kraken, and Sagittarius accretion events respectively. An NSC would not form from a semi-isolated gas cloud as envisioned in the GCZCSE model, but would rather have a much deeper gravitational potential well and potentially new infalling material throughout its life, which could result in a more prolonged and complex star formation history, as is observed for these objects. Therefore, although the GCZCSE model is able to produce clusters with sufficiently large values of $\sigma_0$, that is not the best lens through which to view them
for a number of reasons. First, GCZCSE assumes that the protocluster gas cloud is initially well-mixed and so the only source of abundance variations is from self enrichment, but if infalling gas with a different composition contributed to some of the stars, different stars would effectively have different initial metallicities, which would increase the dispersion. Second, the amount of dispersion depends on the amount of ejecta retained, which depends heavily on the depth of the potential well. If the potential well was significantly deeper than inferred from the mass of stars, or if there was other gas in the vicinity that provided pressure that helped contain the supernova ejecta from escaping, then the amount of self enrichment, and therefore \feh\ spread, would be larger.
Therefore, the derivation of $\fehinit(\sigma_0)$ in these cases may not be correct. Indeed, all three clusters have $\fehinit(\sigma_0)$ that lie well below the AMRs of their associated progenitors. We therefore consider this to be corroborating evidence that these clusters were NSCs, and caution that their \fehinit\ in Table~\ref{table:all} is an underestimate.

\subsection{Ambiguous Associations}

\subsubsection{E 3}

Kinematically assigned as Main/H99 by \citet{Massari19}, E~3 is much more metal-rich than the H99 sequence but sits squarely within the Main AMR. We therefore agree with \citet{Kruijssen20} that E~3 is a main progenitor cluster.

\subsubsection{FSR 1758}

\citet{Myeong19} identified FSR~1758 as part of Sequoia based on kinematics, while \citet{RomeroColmenares21} used a re-analysis of the kinematics and abundance information to argue that it is part of Gaia-Enceladus. Both Sequoia and Gaia-Enceladus have similar AMRs, at least at the $>11.5$~Gyr ages where they overlap, and so we cannot distinguish between these scenarios based on \fehinit.

\subsubsection{NGC 3201 and NGC 6101}

Like FSR~1758, NGC~3201 and NGC~6101 are consistent with both \citet{Massari19} associations of Sequoia and Gaia-Enceladus.

\subsubsection{NGC 5024 and NGC 5053}

Kinematically assigned as H99 by \citet{Massari19} but to the newly-found Wukong stream by \citet{Yuan20} and \citet{Naidu20}, both GCs lie at a similar location in the AMR that is consistent with both origins.

\subsubsection{NGC 5139 ($\omega$ Cen)}

Assigned to either Gaia-Enceladus or Sequoia by \citet{Massari19}, but firmly as Gaia-Enceladus by \citet{Pfeffer21} based on kinematics. When looking at \fehinit, it lies well below the AMR of both progenitors, likely because it is an NSC as argued above, and so the metallicity does not shed any further light on its association.

\subsubsection{NGC 5634}

Kinematically assigned to either Gaia-Enceladus or the Helmi streams by \citet{Massari19}. Based on \feh, it appears to be consistent with the Helmi streams and too metal-poor for Gaia-Enceladus's AMR. When looking at \fehinit, it is at the bottom but not completely discrepant with Gaia-Enceladus's AMR; however, the only similar cluster is NGC~5139, where $\fehinit(\sigma_0)$ is likely too low because it is an NSC. We therefore prefer the association with the Helmi streams.

\subsubsection{NGC 5904}

Kinematically assigned to either Gaia-Enceladus or the Helmi streams by \citet{Massari19}. It lies well within Gaia-Enceladus's AMR, and is also consistent with the upper part of the H99 AMR, so we cannot distinguish between these origins.

\subsubsection{NGC 6121}

Kinematically assigned as Kraken by \citet{Massari19}, but ambiguous between Kraken and the Main progenitor by \citet{Kruijssen20}. When looking at \feh, its metallicity appears to be too high for Kraken, but when looking at \fehinit, it appears consistent with the upper track of Kraken's AMR. We therefore prefer the interpretation that its progenitor is Kraken, but cannot rule out that it formed in the Main proegnitor.

\subsubsection{NGC 6441}

Kinematically assigned as Kraken by \citet{Massari19} but to the Main progenitor by \citet{Kruijssen20} due to its high \feh. When looking at \fehinit, it still lies at the top of Kraken's AMR but is less of an outlier. We therefore consider both interpretations possible.

\subsubsection{Pal 1}

Kinematically assigned as Main by \citet{Massari19}, \citet{Kruijssen20} conclude it was accreted due to its discrepant location on the AMR. The \fehinit\ analysis also shows it to lie in the accreted part of the AMR, so we assign it to the ``high energy'' group.

\section{Conclusions}\label{sec:conclusions}

We have used the the \citet{Bailin18} clumpy self enrichment model GCZCSE for globular cluster (GC) formation to predict the effects of self enrichment on a GC's metallicity, \feh, and internal iron abundance spread, $\sigma_0$, for a large grid of protocluster masses, \Minit, and initial metallicities, \fehinit. We find that the effects of self enrichment are greater for more massive protoclusters, which had a deeper gravitational potential well that was better able to hold onto supernova ejecta, and for those with lower \fehinit, where the relative importance of the new metals is larger. We find that high mass clusters have a metallicity floor where almost all of the final metallicity of the cluster was produced by self enrichment; in this regime, the final metallicity of the cluster is independent of the initial protocluster metallicity, but the degeneracy is broken by the metallicity spread within the cluster: clusters with lower \fehinit\ have larger values of $\sigma_0$.

The GCZCSE model provides a unique mapping between the initial mass and metallicity of the protocluster and the final (observable) metallicity and metallicity spread that we see today, allowing us to reconstruct the initial protocluster metallicity. We have updated the \citet{Bailin19} catalog of iron abundance spreads in observed GCs, and reconstructed the \fehinit\ predicted by the GCZCSE model for all clusters for which $\sigma_0$ and/or \Minit\ is available. We find that both methods of reconstruction give consistent results, but prefer using $\sigma_0$ when available due to its greater robustness.

Among clusters that do not yet have solid measurements of $\sigma_0$, we suggest NGC~6293, NGC~6287, NGC~6541, and NGC~6779 as potentially interesting future targets to search for iron abundance spreads.

We have reconstructed the age-metallicity relations (AMRs) of a number of accretion events in the Milky Way's history that have been proposed in the literature based on kinematics. These histories should be clearer when looking at the initial metallicity, \fehinit, instead of the current metallicity, \feh. In the cases of E~3, NGC~5634, NGC~6121, and Pal~1 this information helps resolve kinematic ambiguities as to which accretion event they are associated with. However, the corrections are mostly small and so the overall AMRs, and therefore the overall framework of the Milky Way's history that has been built up by studying the AMR of GCs, is robust to the impact of self enrichment. There are, however, hints of substructure within the AMRs of the larger accretion events (particularly Gaia-Enceladus, Kraken, and the Helmi streams) that may point to their progenitors having had their own complex formation histories, a topic we will return to in a forthcoming paper (Bailin, in prep.).

Finally, we note that although the GCZCSE model can successfully explain the relationship between the initial protocluster properties and the final metallicity and iron abundance spread in the majority of Milky Way GCs, it assumes that GCs form as semi-isolated systems whose gravitational potential was dominated by the protocluster gas cloud and which acquired no new material. There are 4 Milky Way clusters where it does a poor job because these assumptions were likely violated:
\begin{enumerate}
 \item Terzan~5 has much too high a value of $\sigma_0$ given its high metallicity to be explainable with the GCZCSE model, and has almost certainly been strongly affected by its location deep in the bulge.
 \item NGC~5139 ($\omega$~Cen), NGC~6273, and NGC~6715 (M54) have been previously identified as nuclear star clusters from the Gaia-Enceladus, Kraken, and Sagittarius accretions, and so formed in the centers of dwarf galaxies with deeper potential wells and complicated interactions with their environment. Although the GCZCSE model can produce clusters with their present-day metallicity and $\sigma_0$, the required values of \fehinit\ lie well below the AMRs of their associated accretion events, implying that these other factors have contributed strongly to their iron abundance spreads.
\end{enumerate}

\acknowledgments
We thank the referee for comments which improved the discussions in the paper.

\software{Astropy \citep{astropy:2013,astropy:2018}, emcee \citep{emcee}, Matplotlib \citep{matplotlib}, SciPy \citep{scipy}, Recongc \citep{recongc}}




\bibliographystyle{aasjournal}
\bibliography{gcsigpred}

\begin{thebibliography}{}
\expandafter\ifx\csname natexlab\endcsname\relax\def\natexlab#1{#1}\fi
\providecommand{\url}[1]{\href{#1}{#1}}
\providecommand{\dodoi}[1]{doi:~\href{http://doi.org/#1}{\nolinkurl{#1}}}
\providecommand{\doeprint}[1]{\href{http://ascl.net/#1}{\nolinkurl{http://ascl.net/#1}}}
\providecommand{\doarXiv}[1]{\href{https://arxiv.org/abs/#1}{\nolinkurl{https://arxiv.org/abs/#1}}}

\bibitem[{{Astropy~Collaboration} {et~al.}(2013){Astropy~Collaboration},
  {Robitaille}, {Tollerud}, {Greenfield}, {Droettboom}, {Bray}, {Aldcroft},
  {Davis}, {Ginsburg}, {Price-Whelan}, {Kerzendorf}, {Conley}, {Crighton},
  {Barbary}, {Muna}, {Ferguson}, {Grollier}, {Parikh}, {Nair}, {Unther},
  {Deil}, {Woillez}, {Conseil}, {Kramer}, {Turner}, {Singer}, {Fox}, {Weaver},
  {Zabalza}, {Edwards}, {Azalee Bostroem}, {Burke}, {Casey}, {Crawford},
  {Dencheva}, {Ely}, {Jenness}, {Labrie}, {Lim}, {Pierfederici}, {Pontzen},
  {Ptak}, {Refsdal}, {Servillat}, \& {Streicher}}]{astropy:2013}
{Astropy~Collaboration}, {Robitaille}, T.~P., {Tollerud}, E.~J., {et~al.} 2013,
  \aap, 558, A33, \dodoi{10.1051/0004-6361/201322068}

\bibitem[{{Bailin}(2018)}]{Bailin18}
{Bailin}, J. 2018, \apj, 863, 99, \dodoi{10.3847/1538-4357/aad178}

\bibitem[{{Bailin}(2019)}]{Bailin19}
---. 2019, \apjs, 245, 5, \dodoi{10.3847/1538-4365/ab4812}

\bibitem[{{Bailin}(2021)}]{recongc}
---. 2021, Recongc: Reconstruct globular cluster initial metallicities using
  the GCZCSE model, v1.0.0,  Zenodo, \dodoi{10.5281/zenodo.5111656}

\bibitem[{{Bailin} \& {Harris}(2009)}]{BH09}
{Bailin}, J., \& {Harris}, W.~E. 2009, \apj, 695, 1082,
  \dodoi{10.1088/0004-637X/695/2/1082}

\bibitem[{{Balbinot} \& {Gieles}(2018)}]{Balbinot18}
{Balbinot}, E., \& {Gieles}, M. 2018, \mnras, 474, 2479,
  \dodoi{10.1093/mnras/stx2708}

\bibitem[{{Bastian} \& {Lardo}(2018)}]{BastianLardo18}
{Bastian}, N., \& {Lardo}, C. 2018, \araa, 56, 83,
  \dodoi{10.1146/annurev-astro-081817-051839}

\bibitem[{{Bekki} \& {Tsujimoto}(2019)}]{Bekki19}
{Bekki}, K., \& {Tsujimoto}, T. 2019, \apj, 886, 121,
  \dodoi{10.3847/1538-4357/ab464d}

\bibitem[{{Bellini} {et~al.}(2017){Bellini}, {Milone}, {Anderson}, {Marino},
  {Piotto}, {van der Marel}, {Bedin}, \& {King}}]{Bellini17}
{Bellini}, A., {Milone}, A.~P., {Anderson}, J., {et~al.} 2017, \apj, 844, 164,
  \dodoi{10.3847/1538-4357/aa7b7e}

\bibitem[{{Carretta} {et~al.}(2011){Carretta}, {Lucatello}, {Gratton},
  {Bragaglia}, \& {D'Orazi}}]{Carretta11}
{Carretta}, E., {Lucatello}, S., {Gratton}, R.~G., {Bragaglia}, A., \&
  {D'Orazi}, V. 2011, \aap, 533, A69, \dodoi{10.1051/0004-6361/201117269}

\bibitem[{{Carretta} {et~al.}(2009){Carretta}, {Bragaglia}, {Gratton},
  {Lucatello}, {Catanzaro}, {Leone}, {Bellazzini}, {Claudi}, {D'Orazi},
  {Momany}, {Ortolani}, {Pancino}, {Piotto}, {Recio-Blanco}, \&
  {Sabbi}}]{Carretta09}
{Carretta}, E., {Bragaglia}, A., {Gratton}, R.~G., {et~al.} 2009, \aap, 505,
  117, \dodoi{10.1051/0004-6361/200912096}

\bibitem[{{Chabrier}(2003)}]{Chabrier03}
{Chabrier}, G. 2003, \pasp, 115, 763, \dodoi{10.1086/376392}

\bibitem[{{Edvardsson} {et~al.}(1993){Edvardsson}, {Andersen}, {Gustafsson},
  {Lambert}, {Nissen}, \& {Tomkin}}]{Edvardsson93}
{Edvardsson}, B., {Andersen}, J., {Gustafsson}, B., {et~al.} 1993, \aap, 500,
  391

\bibitem[{{Ferraro} {et~al.}(2016){Ferraro}, {Massari}, {Dalessandro},
  {Lanzoni}, {Origlia}, {Rich}, \& {Mucciarelli}}]{Ferraro16}
{Ferraro}, F.~R., {Massari}, D., {Dalessandro}, E., {et~al.} 2016, \apj, 828,
  75, \dodoi{10.3847/0004-637X/828/2/75}

\bibitem[{{Forbes} \& {Bridges}(2010)}]{ForbesBridges10}
{Forbes}, D.~A., \& {Bridges}, T. 2010, \mnras, 404, 1203,
  \dodoi{10.1111/j.1365-2966.2010.16373.x}

\bibitem[{{Foreman-Mackey} {et~al.}(2013){Foreman-Mackey}, {Hogg}, {Lang}, \&
  {Goodman}}]{emcee}
{Foreman-Mackey}, D., {Hogg}, D.~W., {Lang}, D., \& {Goodman}, J. 2013, \pasp,
  125, 306, \dodoi{10.1086/670067}

\bibitem[{{Harris}(1996)}]{Harris96}
{Harris}, W.~E. 1996, \aj, 112, 1487, \dodoi{10.1086/118116}

\bibitem[{{Harris} {et~al.}(2006){Harris}, {Whitmore}, {Karakla}, {Oko{\'n}},
  {Baum}, {Hanes}, \& {Kavelaars}}]{Harris06}
{Harris}, W.~E., {Whitmore}, B.~C., {Karakla}, D., {et~al.} 2006, \apj, 636,
  90, \dodoi{10.1086/498058}

\bibitem[{{Haywood} {et~al.}(2013){Haywood}, {Di Matteo}, {Lehnert}, {Katz}, \&
  {G{\'o}mez}}]{Haywood13}
{Haywood}, M., {Di Matteo}, P., {Lehnert}, M.~D., {Katz}, D., \& {G{\'o}mez},
  A. 2013, \aap, 560, A109, \dodoi{10.1051/0004-6361/201321397}

\bibitem[{{Horta} {et~al.}(2020){Horta}, {Schiavon}, {Mackereth}, {Beers},
  {Fern{\'a}ndez-Trincado}, {Frinchaboy}, {Garc{\'\i}a-Hern{\'a}ndez},
  {Geisler}, {Hasselquist}, {J{\"o}nsson}, {Lane}, {Majewski},
  {M{\'e}sz{\'a}ros}, {Bidin}, {Nataf}, {Roman-Lopes}, {Nitschelm},
  {Vargas-Gonz{\'a}lez}, \& {Zasowski}}]{Horta20}
{Horta}, D., {Schiavon}, R.~P., {Mackereth}, J.~T., {et~al.} 2020, \mnras, 493,
  3363, \dodoi{10.1093/mnras/staa478}

\bibitem[{Hunter(2007)}]{matplotlib}
Hunter, J.~D. 2007, Computing in Science \& Engineering, 9, 90,
  \dodoi{10.1109/MCSE.2007.55}

\bibitem[{{Jim{\'e}nez} {et~al.}(2021){Jim{\'e}nez}, {Tenorio-Tagle}, \&
  {Silich}}]{Jimenez21}
{Jim{\'e}nez}, S., {Tenorio-Tagle}, G., \& {Silich}, S. 2021, \mnras, 505,
  4669, \dodoi{10.1093/mnras/stab1645}

\bibitem[{{Johnson} {et~al.}(2017){Johnson}, {Caldwell}, {Rich}, \&
  {Walker}}]{Johnson17-N6229}
{Johnson}, C.~I., {Caldwell}, N., {Rich}, R.~M., \& {Walker}, M.~G. 2017, \aj,
  154, 155, \dodoi{10.3847/1538-3881/aa86ac}

\bibitem[{{J{\"o}nsson} {et~al.}(2020){J{\"o}nsson}, {Holtzman}, {Allende
  Prieto}, {Cunha}, {Garc{\'\i}a-Hern{\'a}ndez}, {Hasselquist}, {Masseron},
  {Osorio}, {Shetrone}, {Smith}, {Stringfellow}, {Bizyaev}, {Edvardsson},
  {Majewski}, {M{\'e}sz{\'a}ros}, {Souto}, {Zamora}, {Beaton}, {Bovy}, {Donor},
  {Pinsonneault}, {Poovelil}, \& {Sobeck}}]{Joensson20}
{J{\"o}nsson}, H., {Holtzman}, J.~A., {Allende Prieto}, C., {et~al.} 2020, \aj,
  160, 120, \dodoi{10.3847/1538-3881/aba592}

\bibitem[{{Kruijssen} {et~al.}(2019{\natexlab{a}}){Kruijssen}, {Pfeffer},
  {Crain}, \& {Bastian}}]{Kruijssen19-emosaics}
{Kruijssen}, J.~M.~D., {Pfeffer}, J.~L., {Crain}, R.~A., \& {Bastian}, N.
  2019{\natexlab{a}}, \mnras, 486, 3134, \dodoi{10.1093/mnras/stz968}

\bibitem[{{Kruijssen} {et~al.}(2019{\natexlab{b}}){Kruijssen}, {Pfeffer},
  {Reina-Campos}, {Crain}, \& {Bastian}}]{Kruijssen19-catalog}
{Kruijssen}, J.~M.~D., {Pfeffer}, J.~L., {Reina-Campos}, M., {Crain}, R.~A., \&
  {Bastian}, N. 2019{\natexlab{b}}, \mnras, 486, 3180,
  \dodoi{10.1093/mnras/sty1609}

\bibitem[{{Kruijssen} {et~al.}(2020){Kruijssen}, {Pfeffer}, {Chevance},
  {Bonaca}, {Trujillo-Gomez}, {Bastian}, {Reina-Campos}, {Crain}, \&
  {Hughes}}]{Kruijssen20}
{Kruijssen}, J.~M.~D., {Pfeffer}, J.~L., {Chevance}, M., {et~al.} 2020, \mnras,
  498, 2472, \dodoi{10.1093/mnras/staa2452}

\bibitem[{{Leaman} {et~al.}(2013){Leaman}, {VandenBerg}, \&
  {Mendel}}]{Leaman13}
{Leaman}, R., {VandenBerg}, D.~A., \& {Mendel}, J.~T. 2013, \mnras, 436, 122,
  \dodoi{10.1093/mnras/stt1540}

\bibitem[{{Li} \& {Gnedin}(2019)}]{Li19}
{Li}, H., \& {Gnedin}, O.~Y. 2019, \mnras, 486, 4030,
  \dodoi{10.1093/mnras/stz1114}

\bibitem[{{Ma} {et~al.}(2020){Ma}, {Grudi{\'c}}, {Quataert}, {Hopkins},
  {Faucher-Gigu{\`e}re}, {Boylan-Kolchin}, {Wetzel}, {Kim}, {Murray}, \&
  {Kere{\v{s}}}}]{Ma20}
{Ma}, X., {Grudi{\'c}}, M.~Y., {Quataert}, E., {et~al.} 2020, \mnras, 493,
  4315, \dodoi{10.1093/mnras/staa527}

\bibitem[{{Mar{\'\i}n-Franch} {et~al.}(2009){Mar{\'\i}n-Franch}, {Aparicio},
  {Piotto}, {Rosenberg}, {Chaboyer}, {Sarajedini}, {Siegel}, {Anderson},
  {Bedin}, {Dotter}, {Hempel}, {King}, {Majewski}, {Milone}, {Paust}, \&
  {Reid}}]{MarinFranch09}
{Mar{\'\i}n-Franch}, A., {Aparicio}, A., {Piotto}, G., {et~al.} 2009, \apj,
  694, 1498, \dodoi{10.1088/0004-637X/694/2/1498}

\bibitem[{{Marino} {et~al.}(2015){Marino}, {Milone}, {Karakas}, {Casagrande},
  {Yong}, {Shingles}, {Da Costa}, {Norris}, {Stetson}, {Lind}, {Asplund},
  {Collet}, {Jerjen}, {Sbordone}, {Aparicio}, \& {Cassisi}}]{Marino15}
{Marino}, A.~F., {Milone}, A.~P., {Karakas}, A.~I., {et~al.} 2015, \mnras, 450,
  815, \dodoi{10.1093/mnras/stv420}

\bibitem[{{Marino} {et~al.}(2019){Marino}, {Milone}, {Sills}, {Yong},
  {Renzini}, {Bedin}, {Cordoni}, {D{\textquoteright}Antona}, {Jerjen},
  {Karakas}, {Lagioia}, {Piotto}, \& {Tailo}}]{Marino19}
{Marino}, A.~F., {Milone}, A.~P., {Sills}, A., {et~al.} 2019, \apj, 887, 91,
  \dodoi{10.3847/1538-4357/ab53d9}

\bibitem[{{Marino} {et~al.}(2021){Marino}, {Milone}, {Renzini}, {Yong},
  {Asplund}, {Da Costa}, {Jerjen}, {Cordoni}, {Carlos}, {Dondoglio}, {Lagioia},
  {Jang}, \& {Tailo}}]{Marino21}
{Marino}, A.~F., {Milone}, A.~P., {Renzini}, A., {et~al.} 2021, \apj,
  submitted, arXiv:2106.15978.
\newblock \doarXiv{2106.15978}

\bibitem[{{Massari} {et~al.}(2019){Massari}, {Koppelman}, \&
  {Helmi}}]{Massari19}
{Massari}, D., {Koppelman}, H.~H., \& {Helmi}, A. 2019, \aap, 630, L4,
  \dodoi{10.1051/0004-6361/201936135}

\bibitem[{{Massari} {et~al.}(2014){Massari}, {Mucciarelli}, {Ferraro},
  {Origlia}, {Rich}, {Lanzoni}, {Dalessandro}, {Valenti}, {Ibata}, {Lovisi},
  {Bellazzini}, \& {Reitzel}}]{Massari14}
{Massari}, D., {Mucciarelli}, A., {Ferraro}, F.~R., {et~al.} 2014, \apj, 795,
  22, \dodoi{10.1088/0004-637X/795/1/22}

\bibitem[{{Masseron} {et~al.}(2019){Masseron}, {Garc{\'\i}a-Hern{\'a}ndez},
  {M{\'e}sz{\'a}ros}, {Zamora}, {Dell'Agli}, {Allende Prieto}, {Edvardsson},
  {Shetrone}, {Plez}, {Fern{\'a}ndez-Trincado}, {Cunha}, {J{\"o}nsson},
  {Geisler}, {Beers}, \& {Cohen}}]{Masseron19}
{Masseron}, T., {Garc{\'\i}a-Hern{\'a}ndez}, D.~A., {M{\'e}sz{\'a}ros}, S.,
  {et~al.} 2019, \aap, 622, A191, \dodoi{10.1051/0004-6361/201834550}

\bibitem[{{McKenzie} \& {Bekki}(2018)}]{McKenzie18}
{McKenzie}, M., \& {Bekki}, K. 2018, \mnras, 479, 3126,
  \dodoi{10.1093/mnras/sty1557}

\bibitem[{{M{\'e}sz{\'a}ros} {et~al.}(2020){M{\'e}sz{\'a}ros}, {Masseron},
  {Garc{\'\i}a-Hern{\'a}ndez}, {Allende Prieto}, {Beers}, {Bizyaev},
  {Chojnowski}, {Cohen}, {Cunha}, {Dell'Agli}, {Ebelke},
  {Fern{\'a}ndez-Trincado}, {Frinchaboy}, {Geisler}, {Hasselquist}, {Hearty},
  {Holtzman}, {Johnson}, {Lane}, {Lacerna}, {Longa-Pe{\~n}a}, {Majewski},
  {Martell}, {Minniti}, {Nataf}, {Nidever}, {Pan}, {Schiavon}, {Shetrone},
  {Smith}, {Sobeck}, {Stringfellow}, {Szigeti}, {Tang}, {Wilson}, \&
  {Zamora}}]{Meszaros20}
{M{\'e}sz{\'a}ros}, S., {Masseron}, T., {Garc{\'\i}a-Hern{\'a}ndez}, D.~A.,
  {et~al.} 2020, \mnras, 492, 1641, \dodoi{10.1093/mnras/stz3496}

\bibitem[{{Mieske} {et~al.}(2006){Mieske}, {Jord{\'a}n}, {C{\^o}t{\'e}},
  {Kissler-Patig}, {Peng}, {Ferrarese}, {Blakeslee}, {Mei}, {Merritt}, {Tonry},
  \& {West}}]{Mieske06}
{Mieske}, S., {Jord{\'a}n}, A., {C{\^o}t{\'e}}, P., {et~al.} 2006, \apj, 653,
  193, \dodoi{10.1086/508986}

\bibitem[{{Milone} {et~al.}(2017){Milone}, {Piotto}, {Renzini}, {Marino},
  {Bedin}, {Vesperini}, {D'Antona}, {Nardiello}, {Anderson}, {King}, {Yong},
  {Bellini}, {Aparicio}, {Barbuy}, {Brown}, {Cassisi}, {Ortolani}, {Salaris},
  {Sarajedini}, \& {van der Marel}}]{Milone17}
{Milone}, A.~P., {Piotto}, G., {Renzini}, A., {et~al.} 2017, \mnras, 464, 3636,
  \dodoi{10.1093/mnras/stw2531}

\bibitem[{{Mu{\~n}oz} {et~al.}(2021){Mu{\~n}oz}, {Geisler}, {Villanova},
  {Sarajedini}, {Frelijj}, {Vargas}, {Monaco}, \& {O'Connell}}]{Munoz21}
{Mu{\~n}oz}, C., {Geisler}, D., {Villanova}, S., {et~al.} 2021, \mnras, in
  press, arXiv:2106.15052.
\newblock \doarXiv{2106.15052}

\bibitem[{{Myeong} {et~al.}(2019){Myeong}, {Vasiliev}, {Iorio}, {Evans}, \&
  {Belokurov}}]{Myeong19}
{Myeong}, G.~C., {Vasiliev}, E., {Iorio}, G., {Evans}, N.~W., \& {Belokurov},
  V. 2019, \mnras, 488, 1235, \dodoi{10.1093/mnras/stz1770}

\bibitem[{{Naidu} {et~al.}(2020){Naidu}, {Conroy}, {Bonaca}, {Johnson}, {Ting},
  {Caldwell}, {Zaritsky}, \& {Cargile}}]{Naidu20}
{Naidu}, R.~P., {Conroy}, C., {Bonaca}, A., {et~al.} 2020, \apj, 901, 48,
  \dodoi{10.3847/1538-4357/abaef4}

\bibitem[{{Origlia} {et~al.}(2011){Origlia}, {Rich}, {Ferraro}, {Lanzoni},
  {Bellazzini}, {Dalessandro}, {Mucciarelli}, {Valenti}, \&
  {Beccari}}]{Origlia11}
{Origlia}, L., {Rich}, R.~M., {Ferraro}, F.~R., {et~al.} 2011, \apjl, 726, L20,
  \dodoi{10.1088/2041-8205/726/2/L20}

\bibitem[{{Origlia} {et~al.}(2019){Origlia}, {Mucciarelli}, {Fiorentino},
  {Ferraro}, {Dalessandro}, {Lanzoni}, {Rich}, {Massari}, {Contreras Ramos}, \&
  {Matsunaga}}]{Origlia19}
{Origlia}, L., {Mucciarelli}, A., {Fiorentino}, G., {et~al.} 2019, \apj, 871,
  114, \dodoi{10.3847/1538-4357/aaf730}

\bibitem[{{Pfeffer} {et~al.}(2018){Pfeffer}, {Kruijssen}, {Crain}, \&
  {Bastian}}]{Pfeffer18}
{Pfeffer}, J., {Kruijssen}, J.~M.~D., {Crain}, R.~A., \& {Bastian}, N. 2018,
  \mnras, 475, 4309, \dodoi{10.1093/mnras/stx3124}

\bibitem[{{Pfeffer} {et~al.}(2021){Pfeffer}, {Lardo}, {Bastian}, {Saracino}, \&
  {Kamann}}]{Pfeffer21}
{Pfeffer}, J., {Lardo}, C., {Bastian}, N., {Saracino}, S., \& {Kamann}, S.
  2021, \mnras, 500, 2514, \dodoi{10.1093/mnras/staa3407}

\bibitem[{Piatti \& Geisler(2012)}]{Piatti12}
Piatti, A.~E., \& Geisler, D. 2012, \aj, 145, 17,
  \dodoi{10.1088/0004-6256/145/1/17}

\bibitem[{{Piotto} {et~al.}(2015){Piotto}, {Milone}, {Bedin}, {Anderson},
  {King}, {Marino}, {Nardiello}, {Aparicio}, {Barbuy}, {Bellini}, {Brown},
  {Cassisi}, {Cool}, {Cunial}, {Dalessandro}, {D'Antona}, {Ferraro}, {Hidalgo},
  {Lanzoni}, {Monelli}, {Ortolani}, {Renzini}, {Salaris}, {Sarajedini}, {van
  der Marel}, {Vesperini}, \& {Zoccali}}]{Piotto15}
{Piotto}, G., {Milone}, A.~P., {Bedin}, L.~R., {et~al.} 2015, \aj, 149, 91,
  \dodoi{10.1088/0004-6256/149/3/91}

\bibitem[{{Price-Whelan} {et~al.}(2018){Price-Whelan}, {Sip{\H{o}}cz},
  {G{\"u}nther}, {Lim}, {Crawford}, {Conseil}, {Shupe}, {Craig}, {Dencheva},
  {Ginsburg}, {VanderPlas}, {Bradley}, {P{\'e}rez-Su{\'a}rez}, {de Val-Borro},
  {Paper Contributors}, {Aldcroft}, {Cruz}, {Robitaille}, {Tollerud},
  {Coordination Committee}, {Ardelean}, {Babej}, {Bach}, {Bachetti}, {Bakanov},
  {Bamford}, {Barentsen}, {Barmby}, {Baumbach}, {Berry}, {Biscani}, {Boquien},
  {Bostroem}, {Bouma}, {Brammer}, {Bray}, {Breytenbach}, {Buddelmeijer},
  {Burke}, {Calderone}, {Cano Rodr{\'\i}guez}, {Cara}, {Cardoso}, {Cheedella},
  {Copin}, {Corrales}, {Crichton}, {D{\textquoteright}Avella}, {Deil},
  {Depagne}, {Dietrich}, {Donath}, {Droettboom}, {Earl}, {Erben}, {Fabbro},
  {Ferreira}, {Finethy}, {Fox}, {Garrison}, {Gibbons}, {Goldstein}, {Gommers},
  {Greco}, {Greenfield}, {Groener}, {Grollier}, {Hagen}, {Hirst}, {Homeier},
  {Horton}, {Hosseinzadeh}, {Hu}, {Hunkeler}, {Ivezi{\'c}}, {Jain}, {Jenness},
  {Kanarek}, {Kendrew}, {Kern}, {Kerzendorf}, {Khvalko}, {King}, {Kirkby},
  {Kulkarni}, {Kumar}, {Lee}, {Lenz}, {Littlefair}, {Ma}, {Macleod},
  {Mastropietro}, {McCully}, {Montagnac}, {Morris}, {Mueller}, {Mumford},
  {Muna}, {Murphy}, {Nelson}, {Nguyen}, {Ninan}, {N{\"o}the}, {Ogaz}, {Oh},
  {Parejko}, {Parley}, {Pascual}, {Patil}, {Patil}, {Plunkett}, {Prochaska},
  {Rastogi}, {Reddy Janga}, {Sabater}, {Sakurikar}, {Seifert}, {Sherbert},
  {Sherwood-Taylor}, {Shih}, {Sick}, {Silbiger}, {Singanamalla}, {Singer},
  {Sladen}, {Sooley}, {Sornarajah}, {Streicher}, {Teuben}, {Thomas},
  {Tremblay}, {Turner}, {Terr{\'o}n}, {van Kerkwijk}, {de la Vega}, {Watkins},
  {Weaver}, {Whitmore}, {Woillez}, {Zabalza}, \& {Contributors}}]{astropy:2018}
{Price-Whelan}, A.~M., {Sip{\H{o}}cz}, B.~M., {G{\"u}nther}, H.~M., {et~al.}
  2018, \aj, 156, 123, \dodoi{10.3847/1538-3881/aabc4f}

\bibitem[{{Renzini}(2013)}]{Renzini13}
{Renzini}, A. 2013, \memsai, 84, 162.
\newblock \doarXiv{1302.0329}

\bibitem[{{Romero-Colmenares} {et~al.}(2021){Romero-Colmenares},
  {Fern{\'a}ndez-Trincado}, {Geisler}, {Souza}, {Villanova}, {Longa-Pe{\~n}a
  Dante Minniti}, {Beers}, {Moni Bidin}, {P{\'e}rez-Villegas}, {Moreno},
  {Garro}, {Baeza}, {Henao}, {Barbuy}, {Alonso-Garc{\'\i}a}, {Cohen}, {Lane},
  \& {Mu{\~n}oz}}]{RomeroColmenares21}
{Romero-Colmenares}, M., {Fern{\'a}ndez-Trincado}, J.~G., {Geisler}, D.,
  {et~al.} 2021, \aap, in press, arXiv:2106.00027.
\newblock \doarXiv{2106.00027}

\bibitem[{{Searle} \& {Zinn}(1978)}]{SearleZinn78}
{Searle}, L., \& {Zinn}, R. 1978, \apj, 225, 357, \dodoi{10.1086/156499}

\bibitem[{{Snaith} {et~al.}(2016){Snaith}, {Bailin}, {Gibson}, {Bell},
  {Stinson}, {Valluri}, {Wadsley}, \& {Couchman}}]{Snaith16}
{Snaith}, O.~N., {Bailin}, J., {Gibson}, B.~K., {et~al.} 2016, \mnras, 456,
  3119, \dodoi{10.1093/mnras/stv2788}

\bibitem[{{Strolger} {et~al.}(2020){Strolger}, {Rodney}, {Pacifici}, {Narayan},
  \& {Graur}}]{Strolger20}
{Strolger}, L.-G., {Rodney}, S.~A., {Pacifici}, C., {Narayan}, G., \& {Graur},
  O. 2020, \apj, 890, 140, \dodoi{10.3847/1538-4357/ab6a97}

\bibitem[{{Tang} {et~al.}(2017){Tang}, {Cohen}, {Geisler}, {Schiavon},
  {Majewski}, {Villanova}, {Carrera}, {Zamora}, {Garcia-Hernandez}, {Shetrone},
  {Frinchaboy}, {Meza}, {Fern{\'a}ndez-Trincado}, {Mu{\~n}oz}, {Lin}, {Lane},
  {Nitschelm}, {Pan}, {Bizyaev}, {Oravetz}, \& {Simmons}}]{Tang17}
{Tang}, B., {Cohen}, R.~E., {Geisler}, D., {et~al.} 2017, \mnras, 465, 19,
  \dodoi{10.1093/mnras/stw2739}

\bibitem[{{Tinsley}(1980)}]{Tinsley80}
{Tinsley}, B.~M. 1980, \fcp, 5, 287

\bibitem[{{Villanova} {et~al.}(2019){Villanova}, {Monaco}, {Geisler},
  {O'Connell}, {Minniti}, {Assmann}, \& {Barb{\'a}}}]{Villanova19}
{Villanova}, S., {Monaco}, L., {Geisler}, D., {et~al.} 2019, \apj, 882, 174,
  \dodoi{10.3847/1538-4357/ab3722}

\bibitem[{{Virtanen} {et~al.}(2020){Virtanen}, {Gommers}, {Oliphant},
  {Haberland}, {Reddy}, {Cournapeau}, {Burovski}, {Peterson}, {Weckesser},
  {Bright}, {van der Walt}, {Brett}, {Wilson}, {Jarrod Millman}, {Mayorov},
  {Nelson}, {Jones}, {Kern}, {Larson}, {Carey}, {Polat}, {Feng}, {Moore}, {Vand
  erPlas}, {Laxalde}, {Perktold}, {Cimrman}, {Henriksen}, {Quintero}, {Harris},
  {Archibald}, {Ribeiro}, {Pedregosa}, {van Mulbregt}, \&
  {Contributors}}]{scipy}
{Virtanen}, P., {Gommers}, R., {Oliphant}, T.~E., {et~al.} 2020, Nature
  Methods, 17, 261, \dodoi{https://doi.org/10.1038/s41592-019-0686-2}

\bibitem[{{Wang} {et~al.}(2016){Wang}, {Primas}, {Charbonnel}, {Van der
  Swaelmen}, {Bono}, {Chantereau}, \& {Zhao}}]{Wang16}
{Wang}, Y., {Primas}, F., {Charbonnel}, C., {et~al.} 2016, \aap, 592, A66,
  \dodoi{10.1051/0004-6361/201628502}

\bibitem[{{Wang} {et~al.}(2017){Wang}, {Primas}, {Charbonnel}, {Van der
  Swaelmen}, {Bono}, {Chantereau}, \& {Zhao}}]{Wang17}
---. 2017, \aap, 607, A135, \dodoi{10.1051/0004-6361/201730976}

\bibitem[{{White} \& {Rees}(1978)}]{WhiteRees78}
{White}, S.~D.~M., \& {Rees}, M.~J. 1978, \mnras, 183, 341,
  \dodoi{10.1093/mnras/183.3.341}

\bibitem[{{Willman} \& {Strader}(2012)}]{WillmanStrader12}
{Willman}, B., \& {Strader}, J. 2012, \aj, 144, 76,
  \dodoi{10.1088/0004-6256/144/3/76}

\bibitem[{{Wirth} {et~al.}(2021){Wirth}, {Jerabkova}, {Yan}, {Kroupa}, {Haas},
  \& {{\v{S}}ubr}}]{Wirth21}
{Wirth}, H., {Jerabkova}, T., {Yan}, Z., {et~al.} 2021, \mnras, 506, 4131,
  \dodoi{10.1093/mnras/stab2011}

\bibitem[{{Yuan} {et~al.}(2020){Yuan}, {Chang}, {Beers}, \& {Huang}}]{Yuan20}
{Yuan}, Z., {Chang}, J., {Beers}, T.~C., \& {Huang}, Y. 2020, \apjl, 898, L37,
  \dodoi{10.3847/2041-8213/aba49f}

\end{thebibliography}



\appendix

\section{Data from Meszaros et al. (2020)}

\subsection{Issues at low signal to noise ratio}\label{ap:M20SN}
When incorporating the \citet{Meszaros20} (\citetalias{Meszaros20}) data with our previous measurements, it became clear that the \citetalias{Meszaros20} data always preferred a larger value of $\sigma_0$ than any other data set. Further investigation revealed that this was because \citetalias{Meszaros20} measurements for spectra with $S/N<200$ have systematically lower values of \feh, sometimes by over 0.4~dex.

\begin{figure}
\gridline{
    \fig{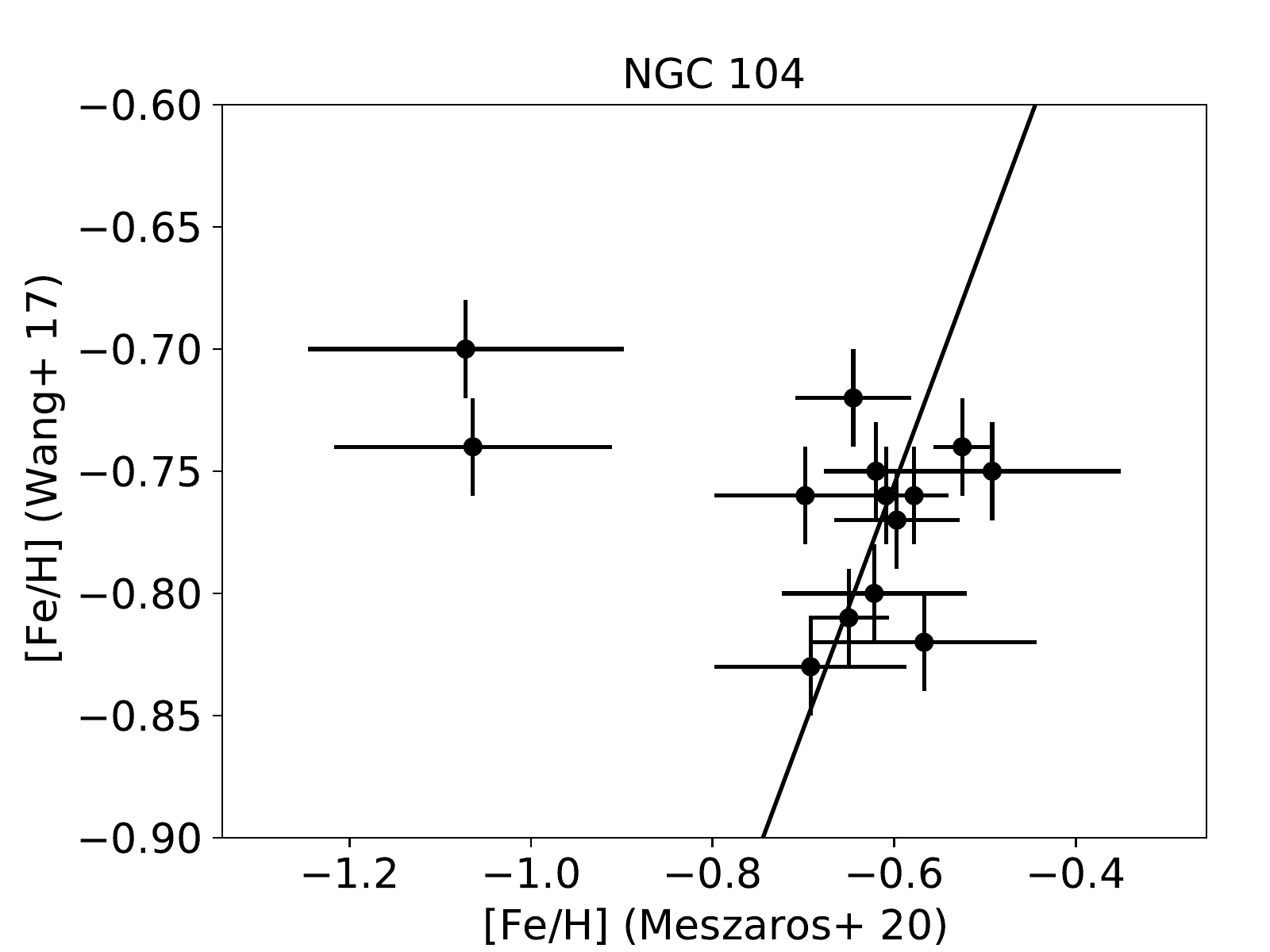}{0.45\textwidth}{(a)}
    \fig{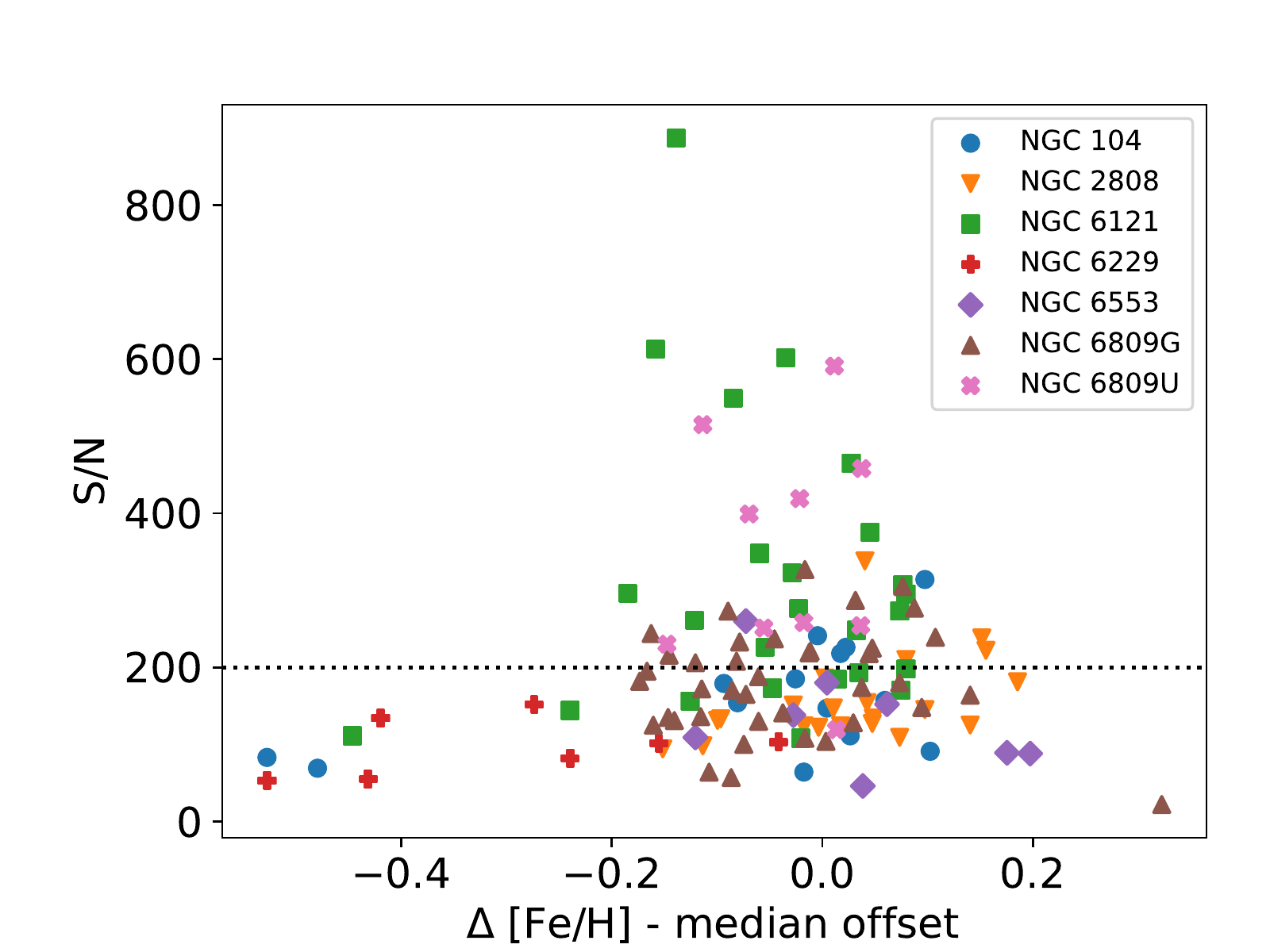}{0.45\textwidth}{(b)}
    }
\gridline{
    \fig{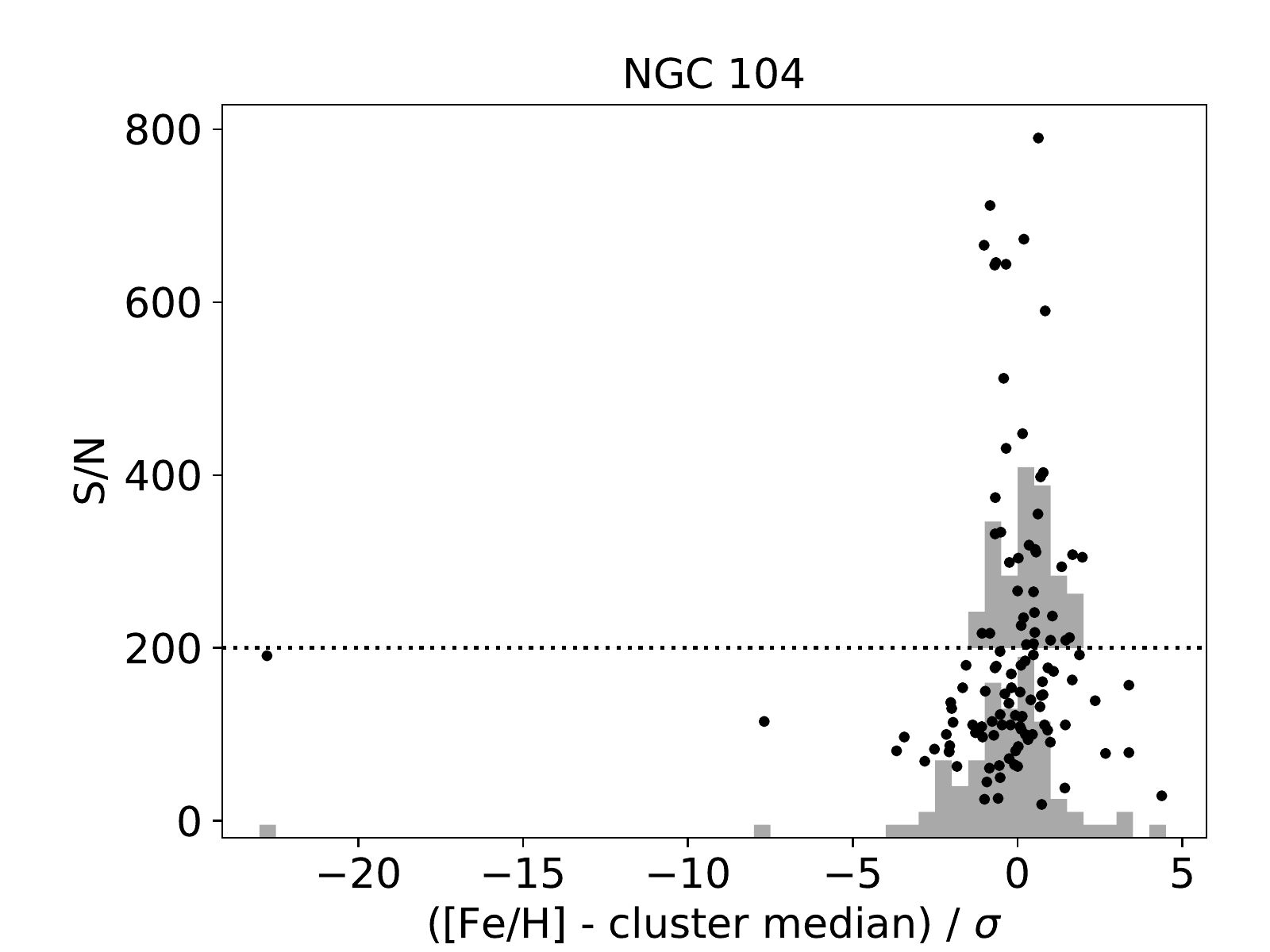}{0.45\textwidth}{(c)}
    \fig{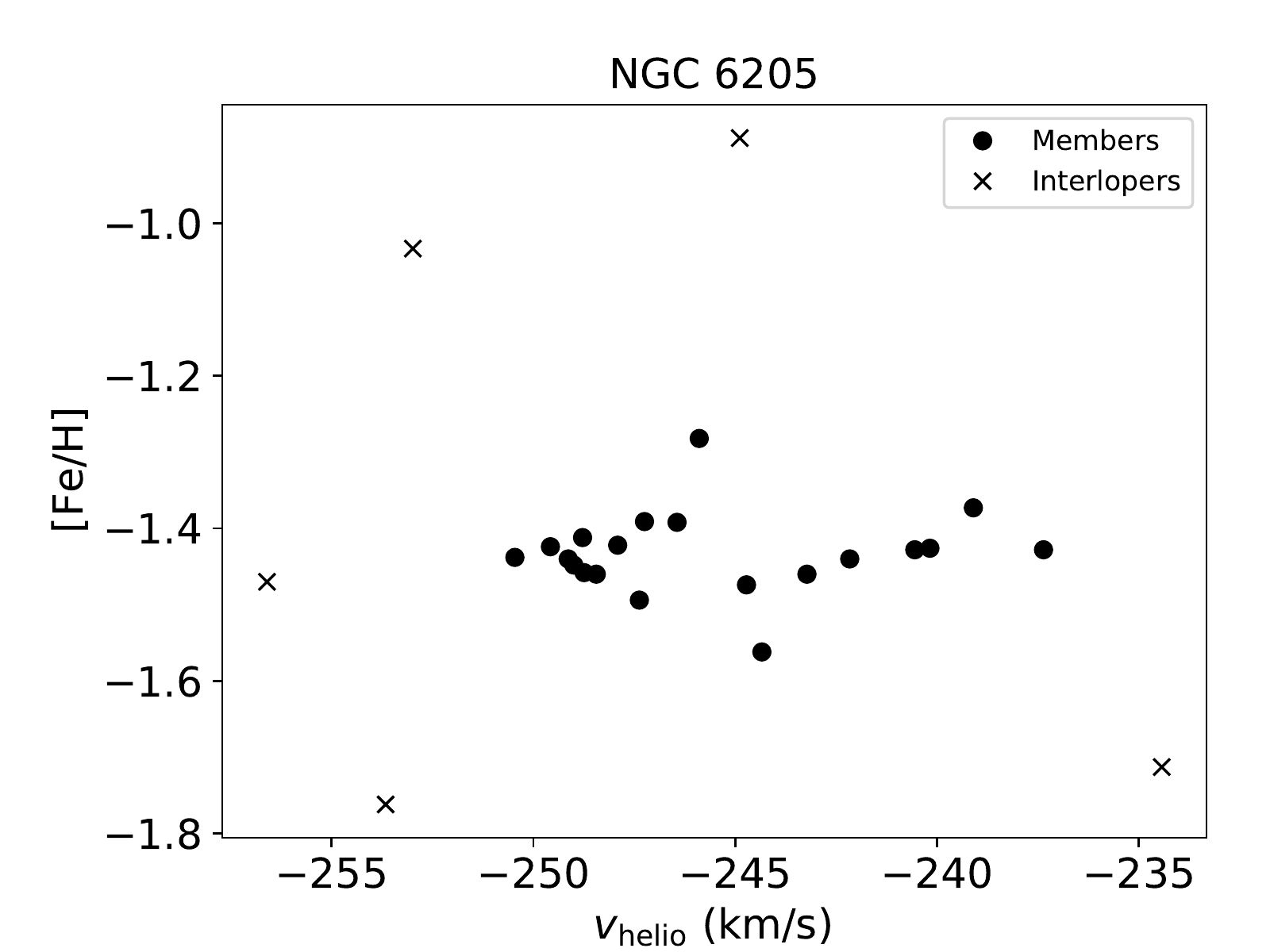}{0.45\textwidth}{(d)}
    }
\caption{\label{fig:N104 FeHcomp}\label{fig:offsetSN}\label{fig:N104 SNoffset}\label{fig:M13-interlopers}
(a) 
[Fe/H] measurements for NGC~104 stars in common between \citet{Meszaros20} and \citet{Wang17}. The black solid line shows the one-to-one relation shifted by the median offset of 0.155~dex. Most points are entirely consistent between the two studies aside from the offset, but two stars have [Fe/H] approximately 0.5~dex lower than expected in \citetalias{Meszaros20}.
(b)
Difference between the \citet{Meszaros20} value of [Fe/H] and the literature value, $\Delta \feh$, versus the \citetalias{Meszaros20} $S/N$, for all matched stars. The median offset has been subtracted out for each cluster. The discrepant measurements with $\Delta \feh < -0.2$ all have $S/N<200$ (dotted black line).
Literature values come from \citet{Wang16,Wang17,Johnson17-N6229,Tang17}; for NGC~6809 the \citet{Wang17} GIRAFFE and UVES data are analyzed separately.
(c)
[Fe/H] measurements for NGC~104 stars in \citet{Meszaros20} relative to the cluster median, divided by the measurement uncertainty, versus the \citetalias{Meszaros20} $S/N$. The dotted black line indicates $S/N=200$. Unlike in panel (a), here we use all \citetalias{Meszaros20} data regardless of whether a matching measurement exists in another data set. The background histograms show the distribution of points for $S/N \ge 200$ (top) and $S/N < 200$ (bottom), which demonstrates that the low $S/N$ data points have a wider distribution and are shifted to the left.
(d)
Metallicity versus radial velocity of all stars listed as belonging to NGC~6205 in \citetalias{Meszaros20} with $S/N>200$. Stars that we have identified as interlopers based on this diagram are marked with an x.
}
\end{figure}

To investigate this, we cross-matched the \citetalias{Meszaros20} data with other studies. There were 6 clusters (7 data sets) where we could easily match individual stars between \citetalias{Meszaros20} and another data set due to the availability of 2MASS identifiers: NGC~104, NGC~2808, NGC~6121, NGC~6809 \citep{Wang16,Wang17}, NGC~6229 \citep{Johnson17-N6229}, and NGC~6553 \citep{Tang17}. We should expect these values to be correlated with each other with a uniform offset, and scatter consistent with the errors. NGC~104 (47~Tuc) is an illustrative example: most stars are consistent, but a few have much lower values of \feh\ -- approximately 0.5~dex -- in \citetalias{Meszaros20} (Figure~\ref{fig:N104 FeHcomp}a).

These discrepant low values are found for stars with lower $S/N$. Figure~\ref{fig:offsetSN}b shows the difference between the \citetalias{Meszaros20} and cross-matched literature value of [Fe/H] (with the cluster median offset subtracted out) versus $S/N$. There is a population of stars with very significant differences ($\Delta\feh < -0.2$, and often $<-0.4$) that are exclusively found at $S/N<200$.

Even without the cross-matched data, internal analysis of all the \citetalias{Meszaros20} data reveals a problem at low $S/N$. Again using NGC~104 as an illustrative example, Figure~\ref{fig:N104 SNoffset}c shows the measurements of \feh\ relative to the cluster median value of $\feh = -0.633$, divided by the measurement uncertainty, versus $S/N$. This should be a vertical swath of points centered at $\sim 0$ with a width that is either uniform, if there is no intrinsic dispersion, or decreasing to lower S/N, if there is intrinsic dispersion. However, below $S/N < 200$ there are two extreme low-\feh\ outliers; moreover, the width of the distribution increases and the center of the distribution moves to lower \feh\ at increasingly low $S/N$. Since $S/N$ is a property of the observation, not of the star, there should not be a physical trend between the metallicity of a star and the $S/N$ of the observation; this trend must be an artefact. All of these effects manifest themselves at $S/N < 200$.

Comparing the top and bottom histograms in Figure~\ref{fig:N104 SNoffset}c makes it clear that including the low $S/N$ points would act to artificially increase the measured dispersion $\sigma_0$. First, the width of the distribution increases at lower $S/N$, which could perhaps be mitigated by recalibrating the uncertainty. However, more insidious is the shift in the central value; adding together even a narrow dataset with an offset central value mimics intrinsic dispersion. We therefore adopt a $S/N=200$ cutoff for \citetalias{Meszaros20} stars to include in our analysis.

\subsection{Interlopers}\label{ap:M20-interlopers}

Examination of the \feh-$v_{\mathrm{helio}}$ distribution of $S/N>200$ stars associated with each cluster sometimes revealed obvious interlopers. The clearest example was NGC~6205 (Figure~\ref{fig:M13-interlopers}d). The list of interlopers we removed in our analysis is given in Table~\ref{tab:M20-interlopers}.

\begin{deluxetable}{ll}
\tablecaption{Interlopers removed from clusters in \citet{Meszaros20}.\label{tab:M20-interlopers}}
\tablehead{ \colhead{Cluster} & \colhead{Star ID}}
\startdata
NGC~288 & 2M00524112-2633271 \\ \hline
NGC~362 & 2M01030778-7049464 \\ \hline
NGC~3201 & 2M10175204-4614066 \\
    & 2M10172586-4626210 \\ \hline
NGC~4590 & 2M12392465-2643331 \\
    & 2M12393987-2647222 \\ \hline
NGC~5272 & 2M13420915+2825401 \\ \hline
NGC~6121 & 2M16224070-2631178 \\ \hline
NGC~6171 & 2M16324155-1302038 \\ \hline
NGC~6205 & 2M16414196+3626518 \\
    & 2M16415030+3624157 \\
    & 2M16412837+3627038 \\
    & 2M16413476+3627596 \\
    & 2M16412450+3629378 \\ \hline
NGC~6218 & 2M16472413-0152522 \\ \hline
NGC~6254 & 2M16570952-0407222 \\ \hline
NGC~6341 & 2M17170043+4305117 \\ \hline
NGC~6809 & 2M19394984-3057240 \\ \hline
NGC~6838 & 2M19534827+1848021 \\
\enddata
\end{deluxetable}



\end{document}